\documentclass[aps,prx,twocolumn,floatfix,superscriptaddress,showpacs,showkeys]{revtex4-1}
\usepackage{epsf,amsmath,amssymb,verbatim,color,multirow,pifont,graphicx,hyperref,cleveref,tabularx,xcolor}

\usepackage{soul}

\crefname{equation}{Eq.}{Eqs.}
\crefname{subsection}{Sec.}{Secs.}
\crefname{figure}{Fig.}{Figs.}
\crefname{table}{Table}{Tables.}

\newcommand{\nbar}{\bar{\nu}}
\newcommand{\nbara}{\bar{\nu}_a}

\newcommand{\psa}{p_{s}}

\newcommand{\asd}{avalanche size distribution}

\begin{document}

\title{Hybrid Phase Transition into an Absorbing State: Percolation and Avalanches}
\author{Deokjae Lee}
\affiliation{CCSS,  CTP and  Department of Physics and Astronomy, Seoul National University, Seoul 08826, Korea}
\author{S. Choi}
\affiliation{CCSS,  CTP and Department of Physics and Astronomy, Seoul National University, Seoul 08826, Korea}
\author{M. Stippinger}
\affiliation{Department of Theoretical Physics, Budapest University of Technology and Economics, Budapest, Hungary}
\author{J. Kert\'esz}
\affiliation{Department of Theoretical Physics, Budapest University of Technology and Economics, Budapest, Hungary}
\affiliation{Center for Network Science, Central European University, Budapest, Hungary}
\author{B. Kahng}
\email{bkahng@snu.ac.kr}
\affiliation{CCSS,  CTP and  Department of Physics and Astronomy, Seoul National University, Seoul 08826, Korea}
\date{\today}

\begin{abstract}
Interdependent networks are more fragile under random attacks than simplex networks, because interlayer dependencies lead to cascading failures and finally to a sudden collapse. This is a hybrid phase transition (HPT), meaning that at the transition point the order parameter has a jump but there are also critical phenomena related to it. Here we study these phenomena on the Erd\H{o}s--R\'enyi and the two dimensional interdependent networks and show that the hybrid percolation transition exhibits two kinds of critical behaviors: divergence of the fluctuations of the order parameter and power-law size distribution of finite avalanches at a transition point. At the transition point global or ``infinite" avalanches occur while the finite ones have a power law size distribution; thus the avalanche statistics also has the nature of a HPT. The exponent $\beta_m$ of the order parameter is $1/2$ under general conditions, while the value of the exponent $\gamma_m$ characterizing the fluctuations of the order parameter depends on the system. The critical behavior of the finite avalanches can be described by another set of exponents, $\beta_a$ and $\gamma_a$. These two critical behaviors are coupled by a scaling law: $1-\beta_m=\gamma_a$.
\end{abstract}


\maketitle

\section{Introduction}
Hybrid phase transitions (HPTs) in complex networks have attracted substantial attention. In these transitions, the order parameter $m(z)$ exhibits behaviors of both first-order and second-order transitions simultaneously as   
\begin{equation}
m(z)=\left\{
\begin{array}{lr}
0 & ~{\rm for}~~  z < z_c, \\
m_0+r(z-z_c)^{\beta_m} & ~{\rm for}~~ z \ge z_c, 
\end{array}
\right.
\label{eq:order}
\end{equation}
where $m_0$ and $r$ are constants and $\beta_m$ is the critical exponent of the order parameter, and $z$ is a control parameter. Examples include the $k$-core percolation~\cite{kcore1,kcore2, kcore_prx}, generalized epidemic spreading~\cite{dodds,janssen,grassberger_epi}, and synchronization \cite{pazo,moreno,mendes_sync}.

Percolation in the cascading failure (CF) model~\cite{buldyrev,baxter,bashan,bianconi,zhou,makse,boccaletti,kivela} on interdependent multi-layer random, Erd\H os-R\'enyi (ER) networks is another example. In this CF model the process is controlled by the mean degree $z$ of the networks \cite{model_comment}. When a node on one layer fails and is deleted, it leads to another failure of the conterpart node in the other layer of the network. Subsequently, links connected to the deleted nodes are also deleted from the networks. This process continues back and forth, always eliminating the possibly separated finite clusters until a giant mutually connected component (MCC) remains or the giant component gets entirely destroyed as a result of the cascades ~\cite{baxter}. As nodes are deleted in such a way, the behavior is similar to that at a second-order phase transition until the transition point $z_c$ is reached from above. Beyond that, as $z$ is further decreased infinitesimally, the percolation order parameter drops suddenly to zero indicating a first-order phase transition. Thus, a HPT occurs at $z=z_c$. This transition may be regarded as a transition to an absorbing state~\cite{absorbing}.  

In the CF model one has to distinguish between clusters and avalanches. Clusters are MCCs~\cite{mcc}. Avalanches consist of MCCs separated from the giant component as a consequence of a triggering removed node and the subsequent cascade \cite{baxter}.
The avalanche sizes depend on the control parameter, the triggering nodes and on the network configurations. 
We call global avalanches with size equal to the order parameter ``infinite", the others are the ``finite'' avalanches.
The size distribution of finite avalanches follows power law at $z_c$~\cite{baxter}, provided the infinite avalanche is discarded. This fact suggests that the avalanche dynamics at $z_c$ exhibits a critical pattern.
The finite MCCs at $z_c$ mostly consist of one or two nodes~\cite{buldyrev,grassberger}, which is in discord with the power-law behavior of the cluster size distribution at a transition point characteristic of the conventional second-order percolation transition \cite{stauffer,christensen}. As the avalanches show critical behavior but the clusters do not, an important challange emerges: How to relate the critical behavior of the order parameter to the avalanche dynamics in a single theoretical framework. Further fundamental questions have been still open, such as how the fluctuations of the order parameter behave at the critical point, whether the scaling relation holds between critical exponents of the order parameter exponent, the susceptibility exponent and the correlation size exponent and whether the hyperscaling relation is valid.
These questions are not limited to the CF model, but are also relevant to other systems undergoing HPT driven by avalanche dynamics, for instance, $k$-core percolation model~\cite{kcore_prx}.

One of the main difficulties in answering those questions has been the need for major computational capacity. Thanks to the efficient algorithm introduced recently by our group \cite{hwang}, we are now able to address those important unsolved problems. In this paper, we report about large scale simulations and analytical results on the CF model of interdependent networks. Based on them we have constructed a theoretical framework connecting the critical behaviors of the order parameter and the avalanche dynamics and have understood the nature of the hybrid percolation transition.  

In this paper we study the interdependent CF model for coupled ER networks and two-dimensional square lattices (2D).
The control parameter for ER (2D) interdependent networks is the average degree $z$ of a node (the fraction $q$ of
original nodes kept in a layer); the order paramater $m$ is the size of the giant mutually connected cluster per node.
To describe the HPT,  we introduce two sets of critical exponents. The set $\{\beta_m, \gamma_m, \nbar_m \}$ is associated with the  order parameter and its related quantities, and the other set $\{\tau_a, \sigma_a, \gamma_a, \nbar_a \}$ is associated with the avalanche size distribution and its related ones. The subscripts $m$ and $a$ refer to the order parameter and avalanche dynamics respectively: The exponent $\beta_m$ is defined by the behavior of the order parameter (Eq.~\ref{eq:order}), and $\gamma_m$ is the exponent of the susceptibility $\chi \equiv N(\langle m^2\rangle -\langle m \rangle^2) \sim (z - z_c)^{-\gamma_m}$ where $N$ is the system size. The exponent $\bar{\nu}_m$ is defined by the finite size scaling behavior of the order parameter: $m - m_0 \sim N^{-\beta_m/\bar{\nu}_m}$ at $z = z_c$. The exponents $\tau_a$, $\sigma_a$, and $\bar{\nu}_a$ characterize the avalanche size distribution $p_s \sim s_a^{-\tau_a} f(s_a / s_a^*)$, where $s_a$ denotes the avalanche size and $f$ is a scaling function. Here $s_a^*$ is the characteristic avalanche size, which behaves as $\sim (z - z_c)^{-1/\sigma_a}$ for $N\to \infty$ and $s_a^* \sim N^{1/\sigma_a \bar{\nu}_a}$ is its finite size scaling at $z_c$. The exponent $\gamma_a$ determines the scaling of the mean size of finite avalanches $\left< s_a \right> \sim (z - z_c)^{-\gamma_a}$. 
One may think naively that the exponents $\bar{\nu}_m$ and $\bar{\nu}_a$ would be the same and $\gamma_m$ and $\gamma_a$ are as well. However, it reveals that those pairs of exponents differ from each other. However, we will show that they are related to each other.

\section{Main results}\label{sec:main_result}

The numerically estimated values of the critical exponents for the ER case are listed in Table I, together with those of the 2D case. For the ER and 2D cases, the hyperscaling relation $\nbar_m=2\beta_m+\gamma_m$ holds even though data collapsing for the 2D case is not as satisfactory as for the ER case. The relation $\sigma_a \nbar_a=\tau_a$ does not hold (Sec.~\ref{sec:num_result_ER} and \ref{sec:num_result_2D}).

The few analytic results related to CF model have been limited so far to locally tree-like graphs where the exponent of the order parameter was found to be $\beta_m=1/2$ and the exponent $\tau_a$ of the avalanche size distribution $p_s$ is $\tau_a=3/2$, with the definition $p_s \sim s^{-\tau_a}$ at $z_c$. We show that $\beta_m=1/2$ is valid not only for tree-like networks but generally for interdependent networks with random dependency links (Sec.~\ref{sec:beta-proof}). Moreover, we also show that the two sets of critical exponents $\{\beta_m, \gamma_m, \nbar_m \}$ and $\{ \tau_a, \sigma_a, \gamma_a, \nbar_a \}$ are not independent of each other. They are coupled through the relation $m(z)+\int_z^{z_0} \langle s_a(z) \rangle \mathrm{d}z =1$, where $z_0$ is the mean degree at the beginning of cascading processes. This leads to $\mathrm{d}m(z)/\mathrm{d}z= \langle s_a(z)\rangle$ and yields $1-\beta_m=\gamma_a$ (Sec.~\ref{sec:sum-rule}). Our numerical values support this relation. 

We classify avalanches in the critical region as finite and infinite avalanches. Infinite avalanche means that the avalanche size is as large as the order parameter. Thus, when it occurs, the GMCC completely collapses, and the system falls into an absorbing state. We find that the mean number of hopping steps denoted as $\langle t \rangle$ between the two layers in avalanche processes depends on the system size $N$ in different ways for the different types of avalanches:   $\langle t \rangle \sim \ln N$ for finite avalanches, and $\sim N^{1/3}$ for infinite avalanches on the ER interdependent network (Sec.~\ref{sec:hop-number}).

\section{Simulation method}\label{method}

The numerical test of the relevant quantities had been a challenging task. 
Recently, however, efficient algorithms have been developed \cite{hwang}, in which the sizes of not only a GMCC  but also other MCCs  can be measured with computational time of $O(N^{1.2})$, as compared to the earlier $O(N^2)$ complexity. (For other algorithms 
see Refs.~\cite{grassberger}) and \cite{herrmann_algo}. Now we can investigate critical properties of the hybrid percolation transition of the CF model thoroughly by measuring various critical exponents including susceptibility and correlation size that were missing in previous studies~\cite{grassberger} for both the ER and two dimensional (2D) lattice interdependent networks. 

\section{Simulation results on the ER interdependent networks}\label{sec:num_result_ER}

We first describe the simulation results on the double-layer ER random networks.
On each layer, an ER network is constructed, with $N$ nodes in both, which are kept fixed. Each node in one layered network has a one-to-one partner node in the other network.
The number of occupied edges $M$ in each layer 
is controlled. The control parameter $z$ is defined as the mean degree 
$z =2M/N$. Using the algorithm~\cite{hwang}, we measure the size of GMCC as a function of $z$. The order parameter $m(z)$ defined as the size of the GMCC  per node, which behaves according to Eq.~(\ref{eq:order}).
To trigger an avalanche and to measure its size, we remove a randomly chosen node in one layer and measure the subsequent decrements of the GMCC size, which sum up to the avalanche size. Then we recover the removed nodes and repeat the above process to obtain a reliable statistics of the avanlanche size distribution for a given point $z$. We simulate $10^4$ network configurations for each system size $N/10^5=4, 16, 64$ and $256$, and $10^3$ configurations for $N/10^5=1024$. We obtain $10^{-4}N$ different avalanche samples for each configuration.

\subsection{Critical behavior of GMCC}
\label{ss:er_crit_order_parameter}

\begin{figure}[h]
\includegraphics[width=.95\linewidth]{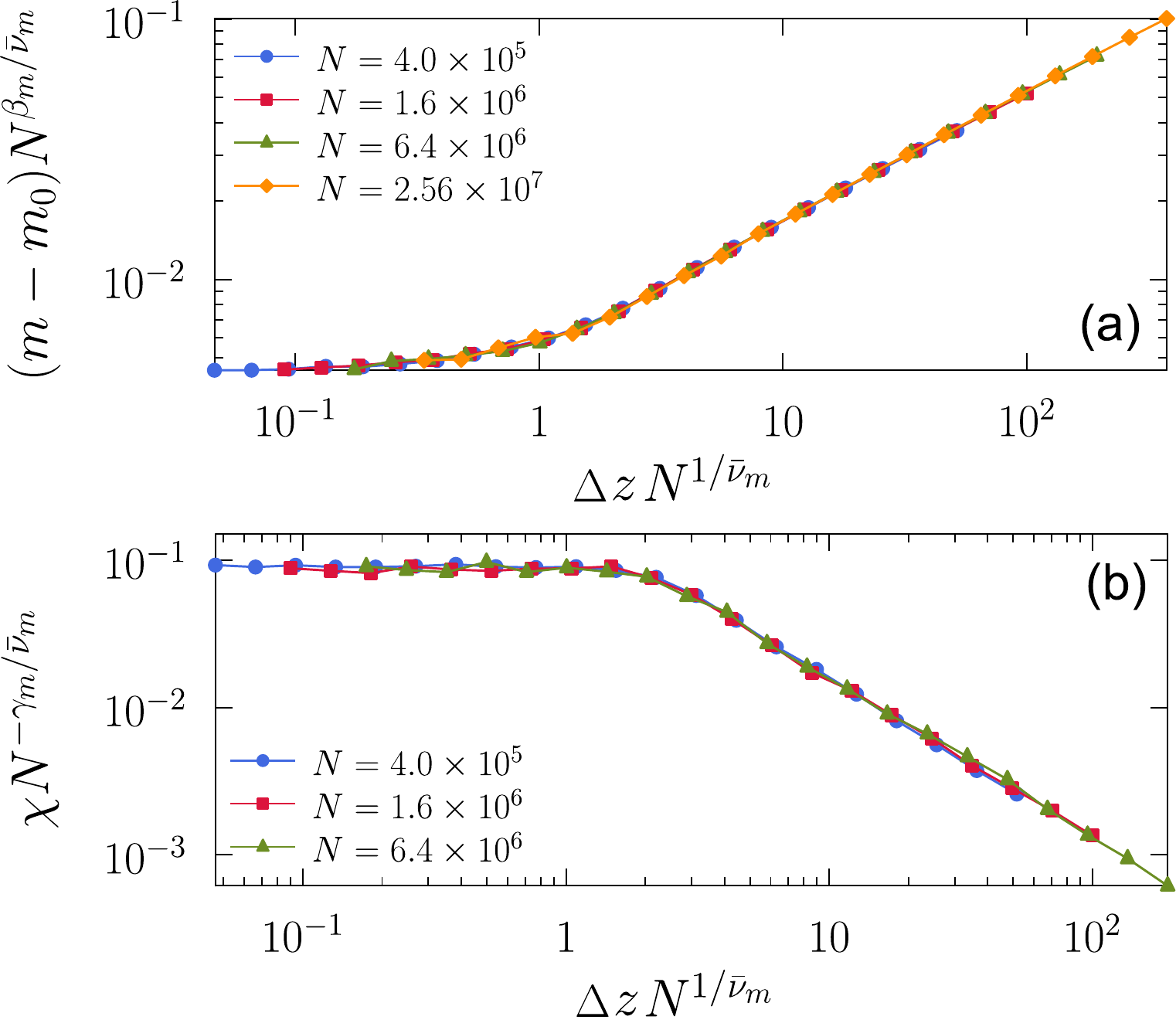}
\caption{
(Color online) (a) Scaling plot of the rescaled order parameter $(m-m_0)N^{\beta_m/\nbar_m}$ vs. $\Delta z N^{1/\nbar_m}$. 
With $\beta_m=0.5$ and $\nbar_m=2.10$ data 
are well collapsed onto a single curve.   
(b) Scaling plot of $(\langle m^2 \rangle - \langle m \rangle^2)N^{1-\gamma_m/\nbar_m}$ 
for different $N$ versus $\Delta z N^{1/\nbar_m}$, where $\gamma_m=1.05$ is used.}
\label{fig:order}
\end{figure}

For the double-layer ER network model, the numerical values of $m_0$ and $z_c$ were obtained in Ref.~\cite{son_grassberger} with high precision as $m_0=0.511700\dots$ and $z_c=2.45540749\dots$. We use these values to evaluate our simulation data. We first check whether Eq.~(\ref{eq:order}) is consistent with the theoretical value $\beta_m=1/2$~\cite{baxter}. In Fig.\ref{fig:order}(a), we plot $(m-m_0)N^{\beta_m/\nbar_m}$ versus $\Delta zN^{1/\nbar_m}$ in scaling form for different system sizes $N$, where $\Delta z\equiv z-z_c(\infty)$. We confirm the exponent to be $\beta_m=0.5\pm 0.01$ from the 
$\Delta z$ region in which the finite-size effect is negligible. Performing finite-size scaling analysis in Fig.~\ref{fig:order}(a), we obtain the correlation size exponent defined as $z_m^*(N)-z_c(\infty)\sim N^{-1/\nbar_m}$ to be $\nbar_m \approx 2.10 \pm 0.02$.

Next, we consider the susceptibility $\chi(z)$ as the fluctuations of the order parameter over the ensemble. This quantity is expected to exhibit critical behavior $\chi \sim (z-z_c)^{-\gamma_m}$ for $z > z_c$. In Fig.~\ref{fig:order}(b),   
we plot a rescaled quantity $(\langle m^2 \rangle-\langle m \rangle^2)N^{1-\gamma_m/\nbar_m}$ versus $\Delta z N^{1/\nbar_m}$. We find that for the 
critical $\Delta z$ region, the data decay in a power-law manner with the exponent $\gamma_m \approx 1.05\pm 0.05$. Moreover, with the choice of $\nbar_m = 2.1$, 
the data are well collapsed onto a single curve. The obtained exponents $\beta_m\approx 0.5\pm 0.01$, $\gamma_m\approx 1.05\pm 0.05$ and $\nbar_m \approx 2.1 \pm 0.02$ satisfy the hyperscaling relation $\nbar_m=2\beta_m+\gamma_m$ reasonably well.

\begin{figure}
\includegraphics[width=.95\linewidth]{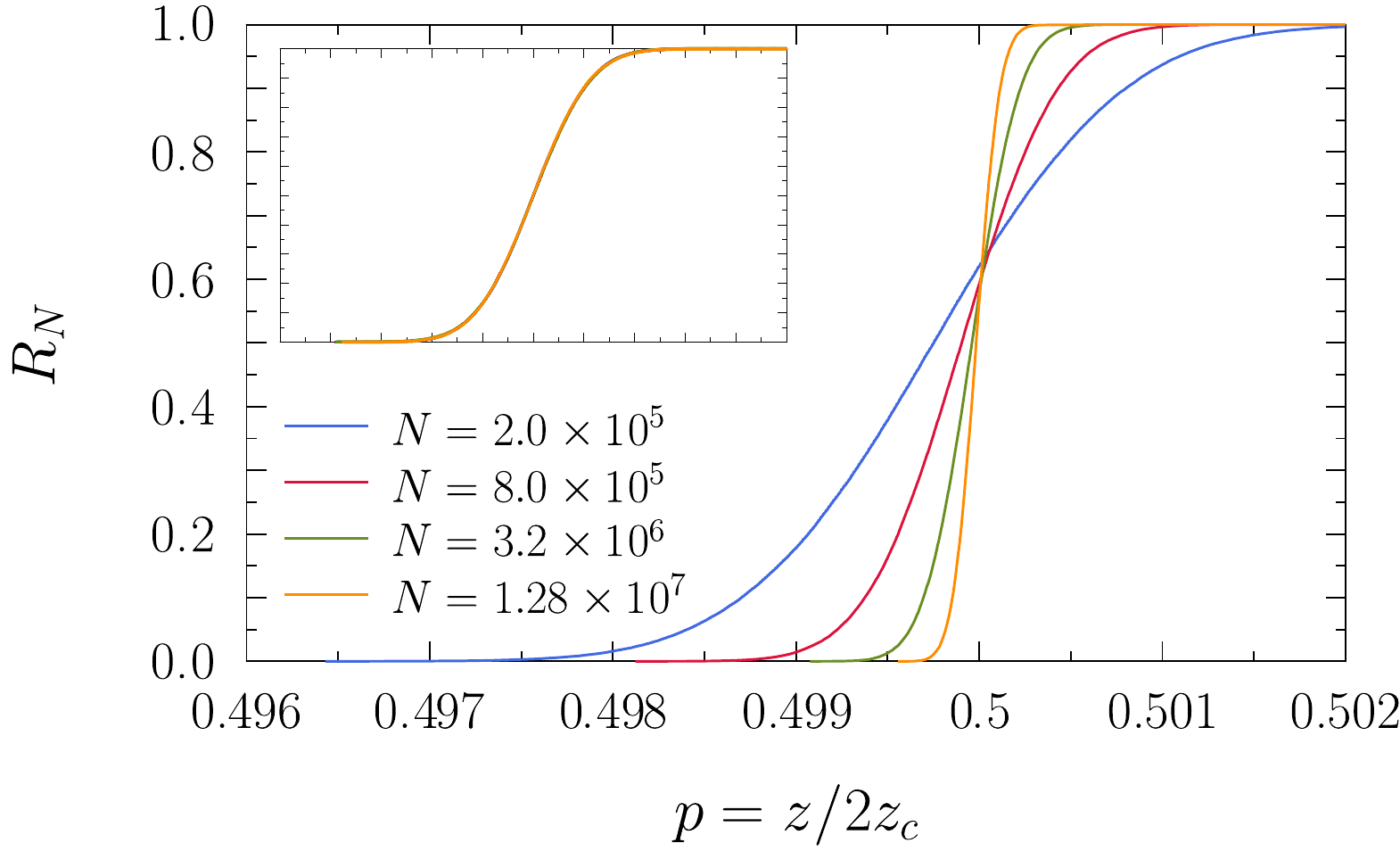}
\caption{
(Color online) The probability $R_N(p)$ that the giant cluster exists. Here $p = z / 2 z_c$ is the node occupation probability of the site percolation in ER networks with mean degree $2 z_c$, which corresponds to the mean degree $z$ of the bond percolation in ER networks. The critical point $p = 0.5$ corresponds to $z = z_c$ in this convention. The inset is the plot of $R_N$ vs.\ $(z - z_c(N)) N^{1/2}$. This data collapse requires $z_c(N)$ instead of $z_c(\infty)$. 
}
\label{fig:RN}
\end{figure}

We also study the probability to contain nonzero GMCC  at a certain point $z$, denoted as $R_N(z)$~\cite{grassberger}. We find that $R_N$ approaches a step function in a form that scales as $R_N([z-z_c(N)]N^{1/2})$ ~(see Fig \ref{fig:RN}). Thus, the slope $\mathrm{d}R_N(z)/\mathrm{d}z$ exhibits a peak at $z_c(N)$, where its value increases as $N^{1/2}$. This means, the probability that the collapse of GMCC  occurs at $z_c(N)$ increases with the rate $N^{1/2}$. Finite size scaling theory suggests the interpretation that $\nbar_m=2$, which is compatible with the result we obtained earlier from Fig.~\ref{fig:order}. One can introduce the order parameter $S(z)$ averaged over all configurations as $S(z)=m(z)R_N(z)$ behaves similarly to the one obtained previously in Fig.~1 of Ref.~\cite{son_grassberger}. 

\begin{figure}
\includegraphics[width=.95\linewidth]{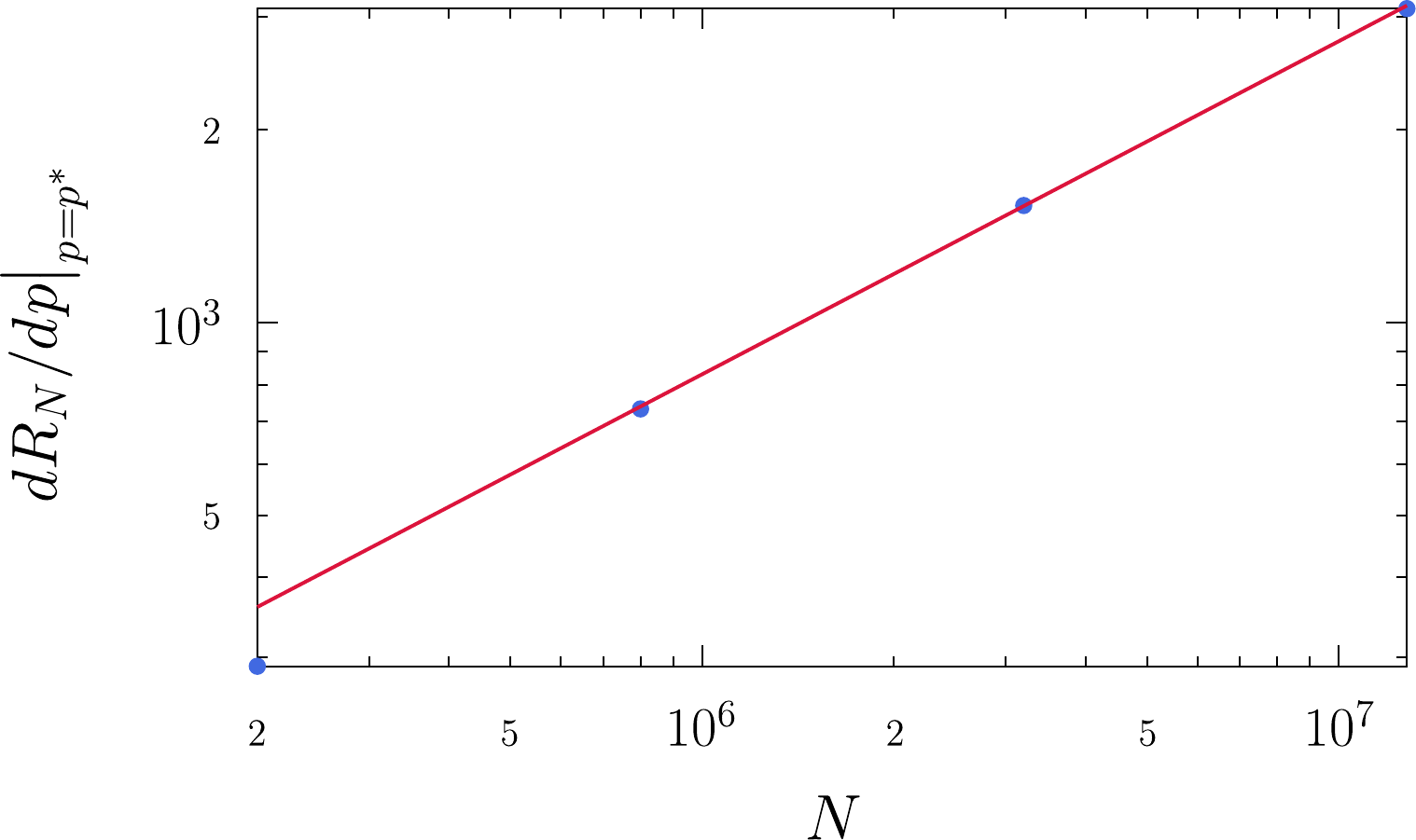}
\caption{
(Color online) The slope of $R_N(p)$ at the fixed point $p^*$, for which $p^* = R_N(p^*)$, as a function of the system size $N$. We measure the slope of the right three data points using the least-square-fit method to be $0.51 \pm 0.02$. Thus, $\nbar_m \approx 1.96\pm 0.07$. Solid line is a guideline with a slope $0.51$
}
\label{fig:Rslope}
\end{figure}

The probability $R_N(z)$ is the basic quantity for large cell renormalization group transformation in percolation theory~\cite{RSK,herrmann_rg}. To proceed, we rescale the control parameter as $p=z/z_0$, where $z_0$ is the mean degree at the beginning of the cascading processes and taken as $z_0=2z_c(\infty)$ for convenience; then $1-p$ is the fraction of nodes removed. Let us define ${\tilde p}=R_N(p)$, where $\tilde p$ can be interpreted as the probability that a node is occupied in a coarse-grained system scaled by $N$.  Using the renormalization group idea, once we find the fixed point $p^*(N)$ satisfying $p^*=R_N(p^*)$ and take the slope $\lambda=\mathrm{d}R_N(p)/\mathrm{d}p$ at $p^*(N)$. Then, we can obtain $\nbar_m=\ln N/\ln \lambda$. Numerically we obtain that $\lambda\sim N^{0.51\pm 0.02}$ and thus $\nbar_m$ is obtained to be $\nbar_m\approx 1.96 \pm 0.07$~(Fig. \ref{fig:Rslope}). This value is close to the one previously obtained by data collapse method.

\begin{figure}[h!]
\includegraphics[width=.95\linewidth]{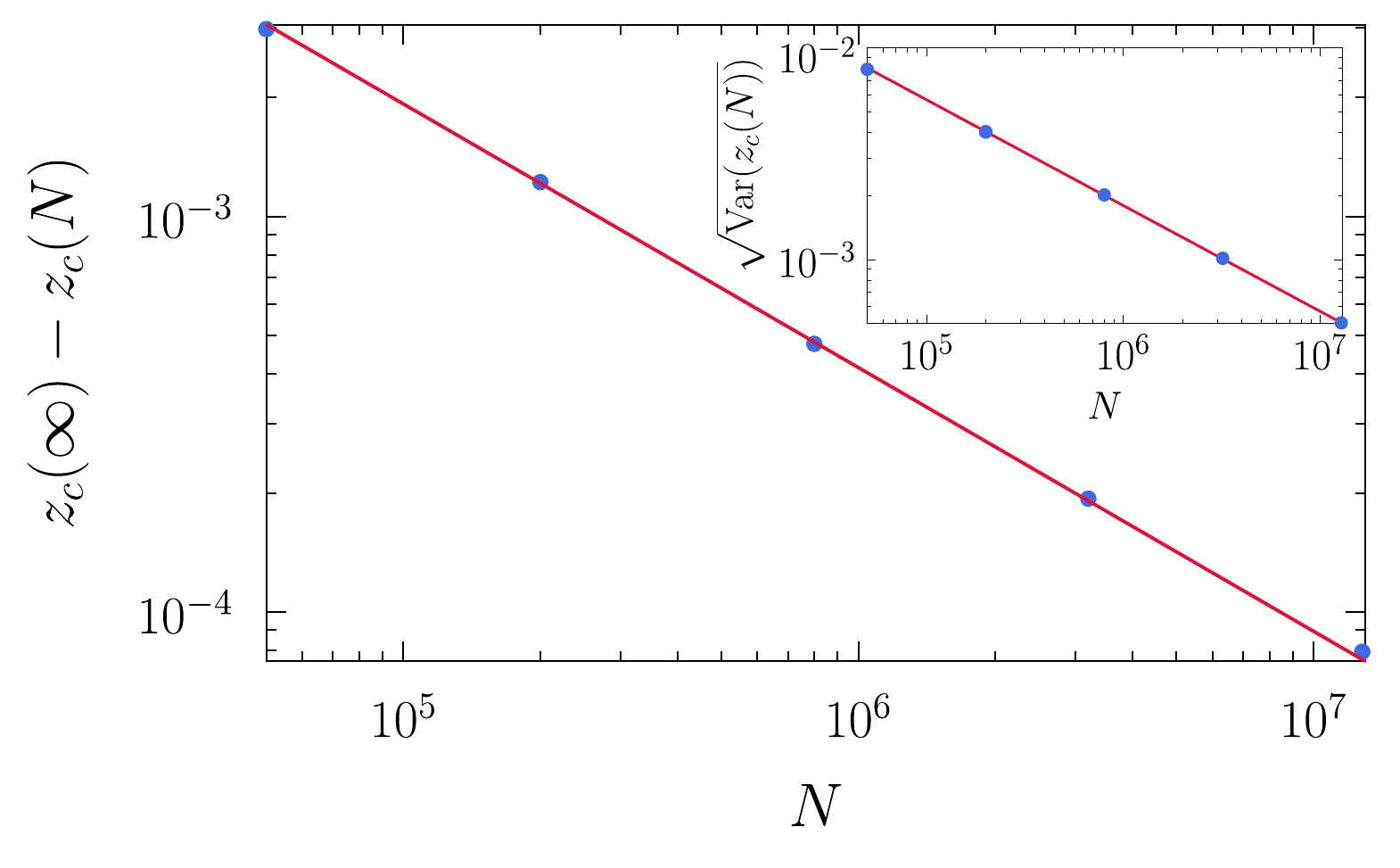}
\caption{(Color online) $z_c(N)$ is the mean position of the order parameter discontinuity when the system size is $N$. It approaches to $z_c(\infty) = 2.45540749\cdots$ as $N$ increases, and the difference scales as $N^{-1/1.5}$. The inset is the standard deviation of $z_c(N)$, which decreases as $N^{-0.5}$.
}
\label{fig:fssfig}
\end{figure}

Interestingly, we measure $p_c-p^*(N)\sim N^{-1/1.5}$ yielding $z_c(\infty)-z^*(N)\sim N^{-1/1.5}$. Similarly, from direct simulations we obtain $z_c(\infty)-z_c(N)\sim N^{-1/1.5}$, where $z_c(N)$ is the average finite size transition point (Fig \ref{fig:fssfig}). In a conventional second-order transition, we would expect that these quantities scale with $N$ as $N^{-1/\nbar_m}$. The difference to $\nbar_m\approx 2$ indicates either an additional diverging scale or extraordinarily large corrections.
But as we have seen previously, the standard definition of the exponent yields $\nbar_m\approx 2$. This is confirmed by the inset of \cref{fig:fssfig} which shows $\sqrt{\mathrm{Var}(z_c(N))}\sim N^{-0.5}$, where $\mathrm{Var}$ stands for the variance. We conclude $\nbar_m\approx 2$ which is also consistent with the value we obtained using the renormalization group transformation eigenvalue. 

\subsection{Critical behavior of avalanche dynamics}

\begin{figure}[t]
\includegraphics[width=.95\linewidth]{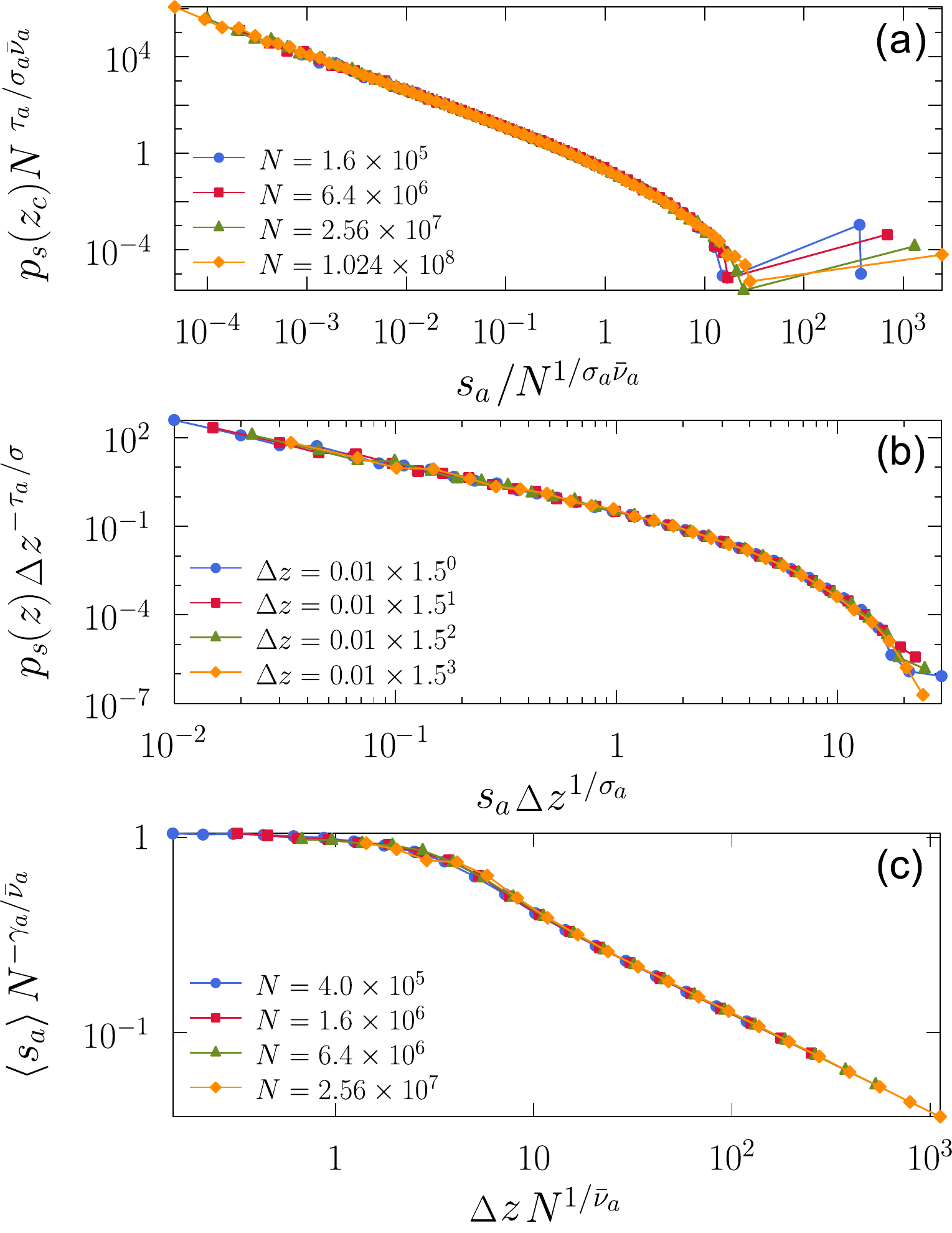}
\caption{(Color online) 
(a) Scaling plot of $p_s(z_c)N^{\tau_a/\sigma_a \nbara}$ vs $s_a/N^{1/\sigma_a\nbara}$ for different system sizes, with $\tau_a=1.5$ and $\sigma_a \nbara\approx 1.85$. 
Note that infinite avalanche sizes for different $N$ do not collapse onto a single dot, because they depend on $N$ as $s_{a,\infty}^* \sim N$. 
(b) Scaling plot of $p_s(z)\Delta z^{-\tau_a/\sigma_a}$ vs $s_a\Delta z^{1/\sigma_a}$ for different $\Delta z$ but a fixed system size $N = 2.56\times10^7$, with $\tau_a=1.5$ and $\sigma_a \approx1.01$.
(c) Scaling plot of $\langle s_a \rangle N^{-\gamma_a/\nbara}$ vs $\Delta z N^{1/\nbara}$ for different system sizes and $\gamma_a=0.5$.
}
\label{fig:aval}
\end{figure}

To characterize the avalanche processes, we count the avalanche size defined as the number of nodes removed in each layer during the cascading processes, denoted as $s_a(z)$. The distribution of those avalanche sizes collected from different triggering nodes and configurations is denoted as $\psa(z)$. In Ref.~\cite{baxter} 
analytically $\psa(z_c) \sim s_a^{-\tau_a}$ with $\tau_a=3/2$ was obtained for locally tree like graphs. We confirm this exponent value 
in Fig.~\ref{fig:aval}(a). 
Avalanches in the region $z < z_a^*(N)$ need to be classified as finite or infinite avalanches;
the latter locate separately in Fig.~\ref{fig:aval}(a). Infinite avalanche means the avalanche size is as large as $m(z)$, i.e., the GMCC  completely collapses, and the system falls into an absorbing state. The infinite avalanche begins to appear at $z=z_a^*(N)$. Fig.~\ref{fig:aval}(a) shows the scaling behavior of the \asd in form of $\psa N^{\tau_a/\sigma_a \nbara}$ versus  $s_aN^{-1/\sigma_a \nbara}$ at $z_c$. The data from different system sizes are well collapsed onto a single curve by the choices of $\tau_a=3/2$ and $\sigma_a\nbara\approx 1.85$. This result suggests that there exists a characteristic size  $s_a^*\sim N^{1/\sigma_a \nbara}$ with   $\sigma_a\nbara\approx 1.85 \pm 0.02$ for finite avalanches. These values indicate that the hyperscaling relation $\sigma_a\nbara=\tau_a$ does not hold for the avalanche dynamics. 
For infinite avalanches, $s_{a,\infty}^*\sim O(N)$. 

For $z > z_c$, we examine the avalanche size distribution versus $s_a$ for different $\Delta z$, and find that it behaves as $\psa \sim s_a^{-\tau_a} f(s_a/s_a^*)$ where $f$ is a scaling function. Following conventional percolation theory~\cite{stauffer}, we assume $s_a^{*}\sim \Delta z^{-1/\sigma_a}$. The exponent $\sigma_a$ is obtained from the scaling plot of $\psa(z)\Delta z^{-\tau_a/\sigma_a}$ vs $s_a\Delta z^{1/\sigma_a}$  in Fig.~\ref{fig:aval}(b). The data are well collapsed with $\sigma_a\approx 1.0$, leading to $\nbar_a\approx 1.85$. This is different from $\bar{\nu}_m$ and indicates that there exists another divergent scale.

\begin{figure}[t!]
\includegraphics[width=.95\linewidth]{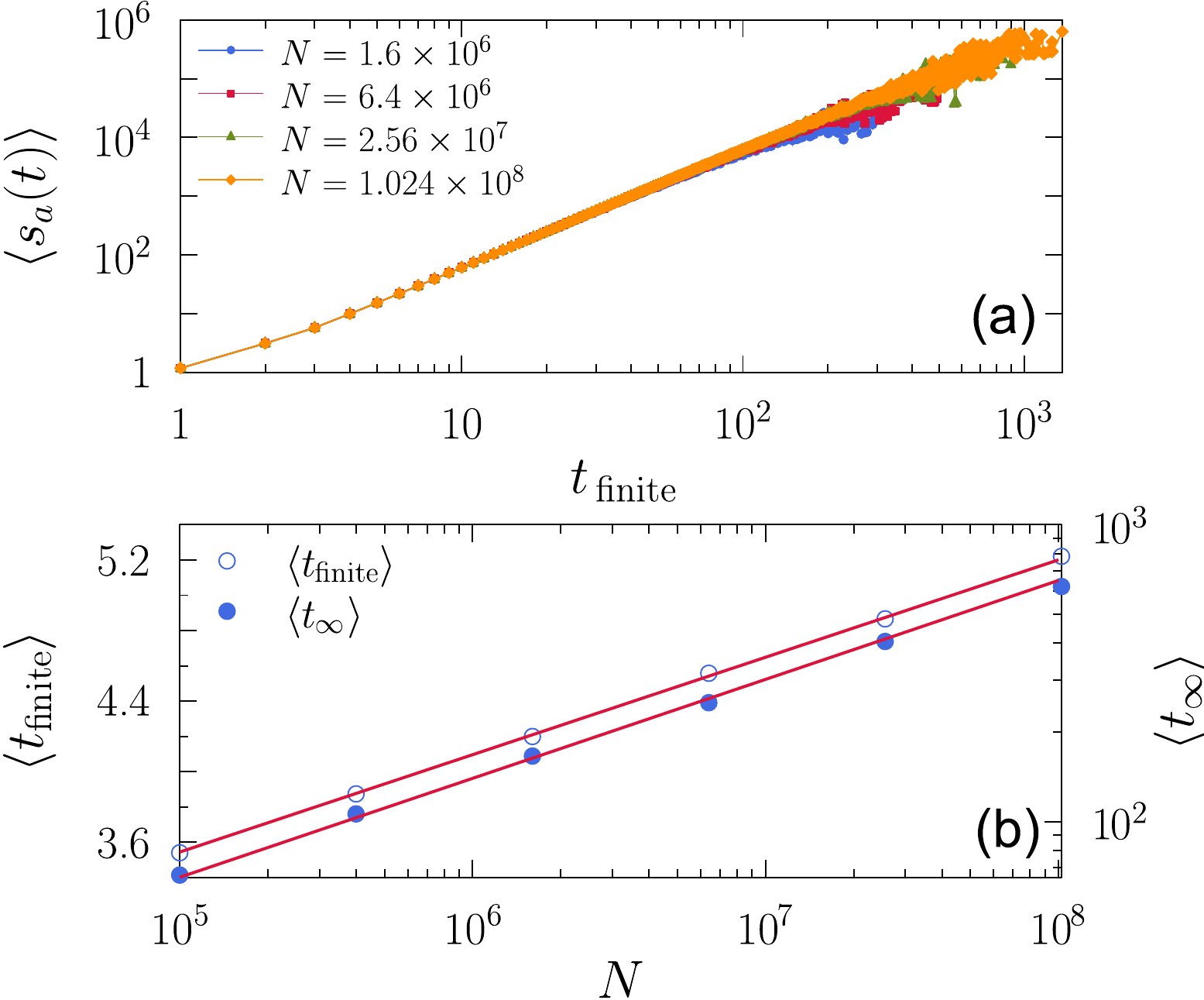}
\caption{(Color online) 
(a) Plot of $\langle s_a(t) \rangle$ as a function of $t$ at $z_c$ for finite avalanches, showing
$\langle s_a(t) \rangle \sim t^{2.0\pm 0.01}$. (b) The plot of $\langle t_{\rm finite} \rangle$ of finite avalanches vs. $N$ at $z_c$ on semi-logarithmic scale (left axis). Plot of $\langle t_{\infty} \rangle$ of infinite 
avalanches as a function of $N$ on double-logarithmic scale (right axis). The guide line has a slope of 
$1/3$.
}
\label{fig:hop}
\end{figure}

We examine the mean avalanche size $\langle s_a\rangle\equiv \sum_{s_a=1}^{\prime} s_a\psa(z)\sim (\Delta z)^{-\gamma_a}$, where the prime indicates
summation 
over finite avalanches. It follows that $\gamma_a=(2-\tau_a)/\sigma_a$~\cite{stauffer}. Thus, $\gamma_a=0.5$ is expected. Our simulation 
confirms this 
value in the large $\Delta z$ region (Fig.~\ref{fig:aval}(c)). Data from different system sizes are well collapsed in the plot of $\langle s_a \rangle N^{-\gamma_a/\nbara}$ vs $\Delta z N^{1/\nbara}$ with $\gamma_a=0.5$ and $\nbar_a=1.85$. This means that there exists crossover points $z_a^*(N)$ such that $z_a^*(N)-z_c\sim N^{-1/\nbar_a}$ in finite systems.
In the thermodynamic limit, $\langle s_a \rangle/N$ is equal to $0$ for $z < z_c$, $s_0$ for $z = z_c$ and $w(z-z_c)^{-\gamma_a}$ for $z > z_c$ where $s_0$ is constant and $w\sim O(N^{-1})$. 
This result shows that the avalanche statistics also exhibits HPT. 

\subsection{Statistics of the number of hops}
\label{sec:hop-number}

When investigating the avalanche dynamics we first focus on {\it finite avalanches}.
Let $t_i(z)$ be the number of hopping steps between the two layers in avalanche processes, when the $i$th node is removed from the GMCC at $z$. $\langle s_a(t) \rangle_i$ is the avalanche size averaged over $i$, that is, the mean number of nodes removed, accumulated up to steps $t$. It is found in Fig.~\ref{fig:hop}(a) that $\langle s_a(t) \rangle \sim t^2$ for finite avalanches, similarly to~\cite{zhou}. 
Using the \asd~$p_s(z)$, we set up the duration time distribution $p_t(z)$ through the relations $p_s \mathrm{d}s =p_t \mathrm{d}t$ and $s_a \sim t^2$ as $p_{t}(z) \sim t^{-2\tau_a+1} f(t^2/(\Delta z)^{-1/\sigma_a})$. 
The mean number of hopping steps for finite avalanches is $\langle t_{\rm finite} \rangle \equiv \sum_{t=1}^{\prime} t p_t(z)$. Because of $\tau_a=3/2$,  $\langle t \rangle \sim -\ln (\Delta z)$ 
for $z > z_c$ and $\langle t_{\rm finite} \rangle \sim \ln N$ at $z=z_c$ (Figs.~\ref{fig:hop}(b) and \ref{fig:avhops}). 
The number of hopping steps of {\it infinite avalanches} 
which can 
appear in the region $z < z_a^*(N)$ lead to $\langle t_{\infty} \rangle \sim N^{1/3}$, 
as shown in Fig.~\ref{fig:hop}(b), in agreement with 
\cite{zhou}. 

\begin{figure}[h]
\includegraphics[width=.95\linewidth]{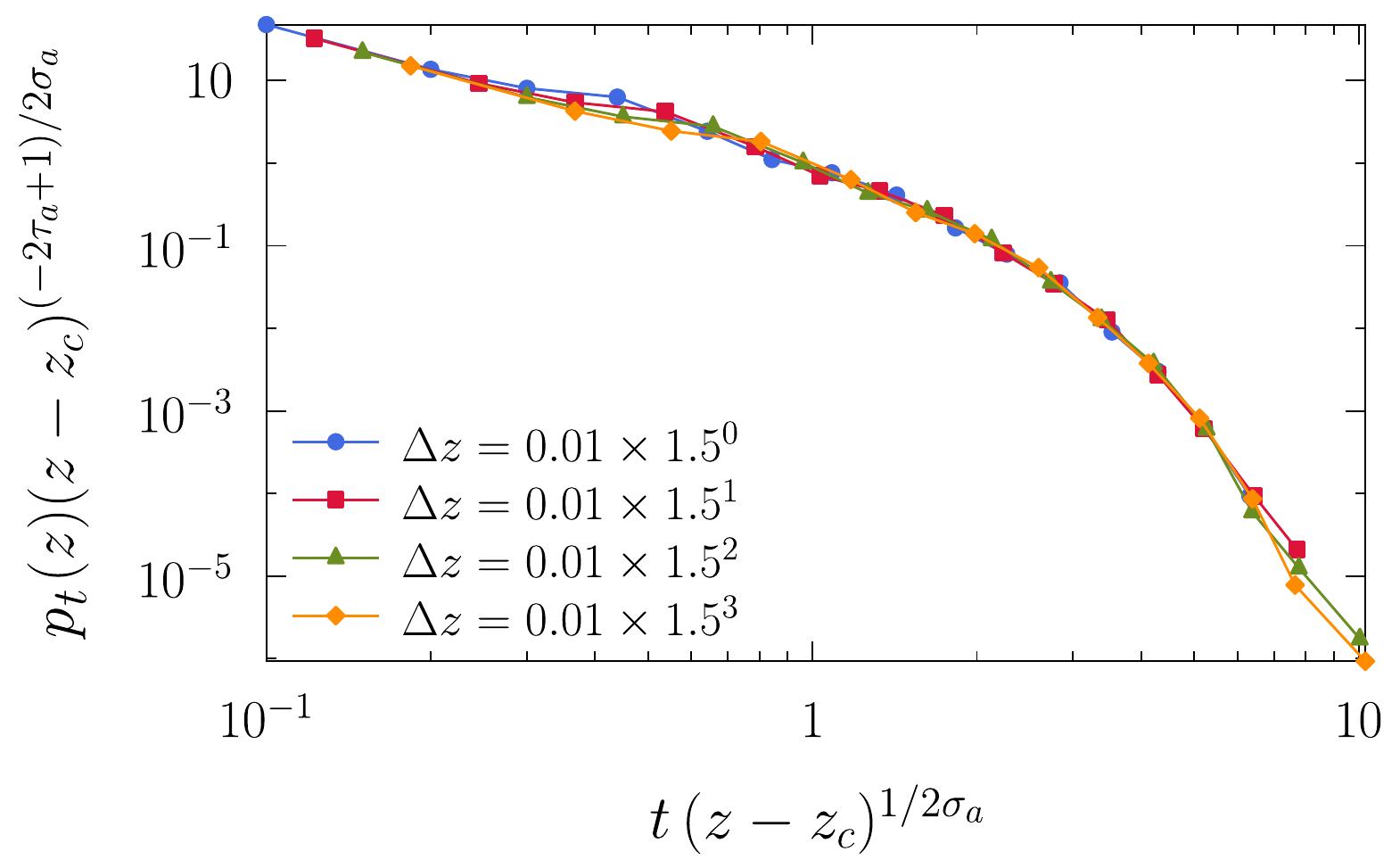}
\caption{(Color online)
The scaling collapse of the distribution $p_t$ of the number of hops $t$ for finite avalanches.
}
\label{fig:hopscaling}
\end{figure}

\begin{figure}
\includegraphics[width=.95\linewidth]{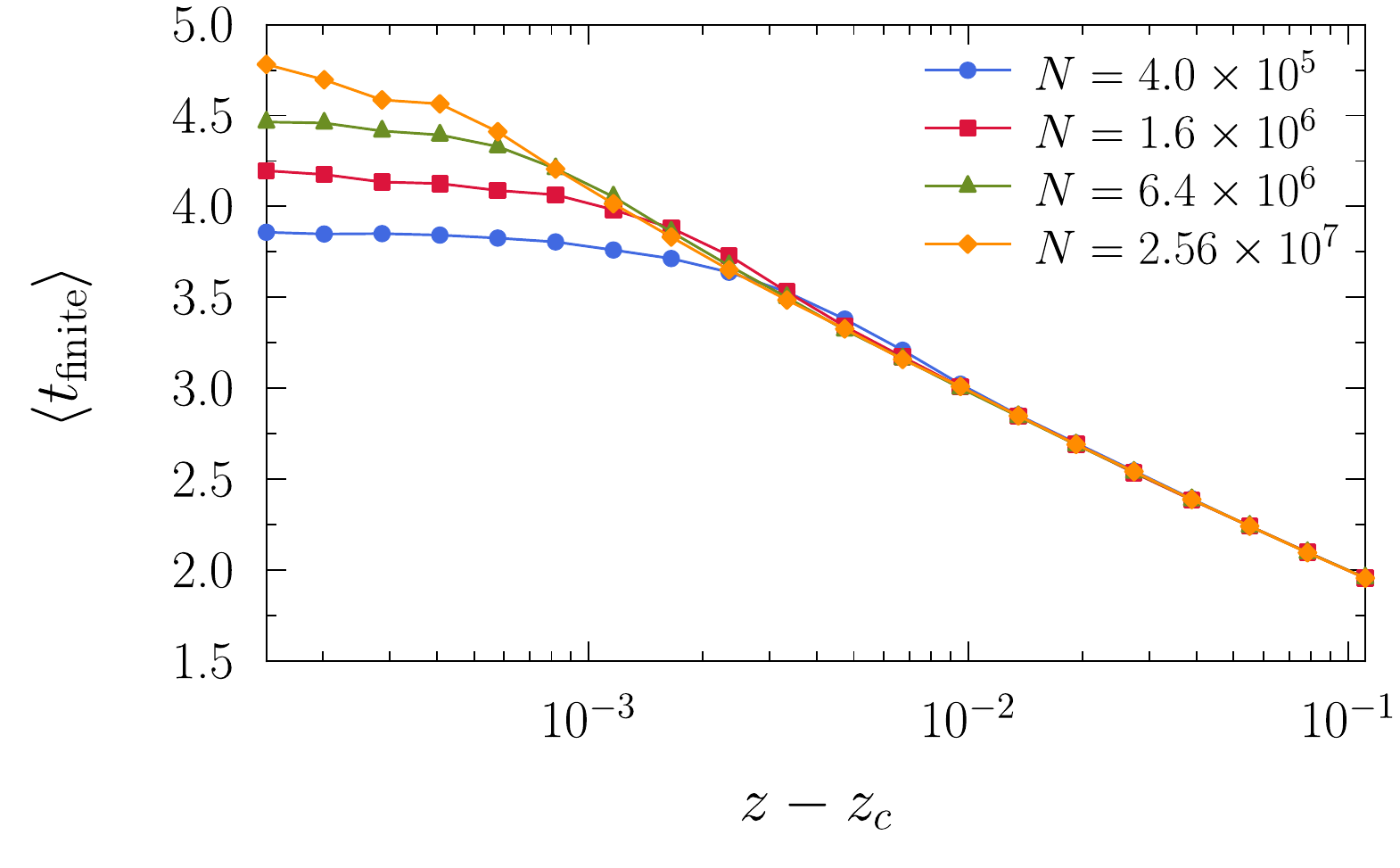}
\caption{(Color online)
The mean number of hops of finite avalanches $\left< t_{\text{finite}} \right>$ as a function of the mean degree $z$.
}
\label{fig:avhops}
\end{figure}

The scaling plot $p_t(z)(z-t_c)^{(-2\tau_a+1)/2\sigma_a}$ vs.\ $t(z-z_c)^{1/2\sigma_a}$ displayed in \cref{fig:hopscaling} proves our hypothesis of $p_t(z)\sim t^{(-2\tau_a+1)}f(-t^2/(\Delta z)^{-1\sigma_a})$. The exactly known special value $\tau_a=3/2$ yields $\langle t_{\mathrm{finite}}\rangle \sim \ln N$ at $z=z_c$ as observed in Fig.~\ref{fig:hop}(b). 

\section{Simulation results on the 2D interdependent networks}\label{sec:num_result_2D}

Let us describe the CF model on two layers of randomly interdependent 2D  networks~\cite{son_grassberger,Li}. At the beginning the layers consist of topologically identical square lattices of size $N=L\times L$ sites with nearest-neighbor connectivity links within each layer. As it was the case for ER networks, the set of nodes in one layer has a random one-to-one correspondence via dependency links with the set of nodes in the other layer. 

The control parameter is defined as the fraction $q$ of original nodes kept in a layer \cite{buldyrev}, analogously to the site percolation problem. Each node shares its fate with its interdependent node on the other layer. The order parameter $m(q)$ is defined as the relative size of the GMCC.

We applied two boundary conditions (BC-s) to the system: periodic and semiperiodic. In the periodic BC the system is on a torus, while in the semiperiodic BC it is on a cylinder, i.e., open in one direction and periodic in the other one. The order of the characteristic parameter values ($q_c(\infty)$, $q_c(N)$ and $q_m^*(N)$) depends on the BC. For the periodic (semiperiodic) BC we have $q_c(N)<q_c(\infty)<q_m^*(N)$ ($q_c(\infty)<q_c(N)<q_m^*(N)$).

The average order parameter $m_c(N)$ before collapse is defined as the smallest nonzero values of the relative size of the giant component averaged over all runs with size $N$. For periodic (semiperiodic) BC we have $m_c(\infty)>m_c(N)$ ($m_c(\infty)<m_c(N)$). The figures are for semiperiodic BC if not indicated otherwise.

\subsection{Critical behavior of GMCC}
\label{ss:2d_crit_order_parameter}

\begin{figure}[h]
\includegraphics[width=.95\linewidth]{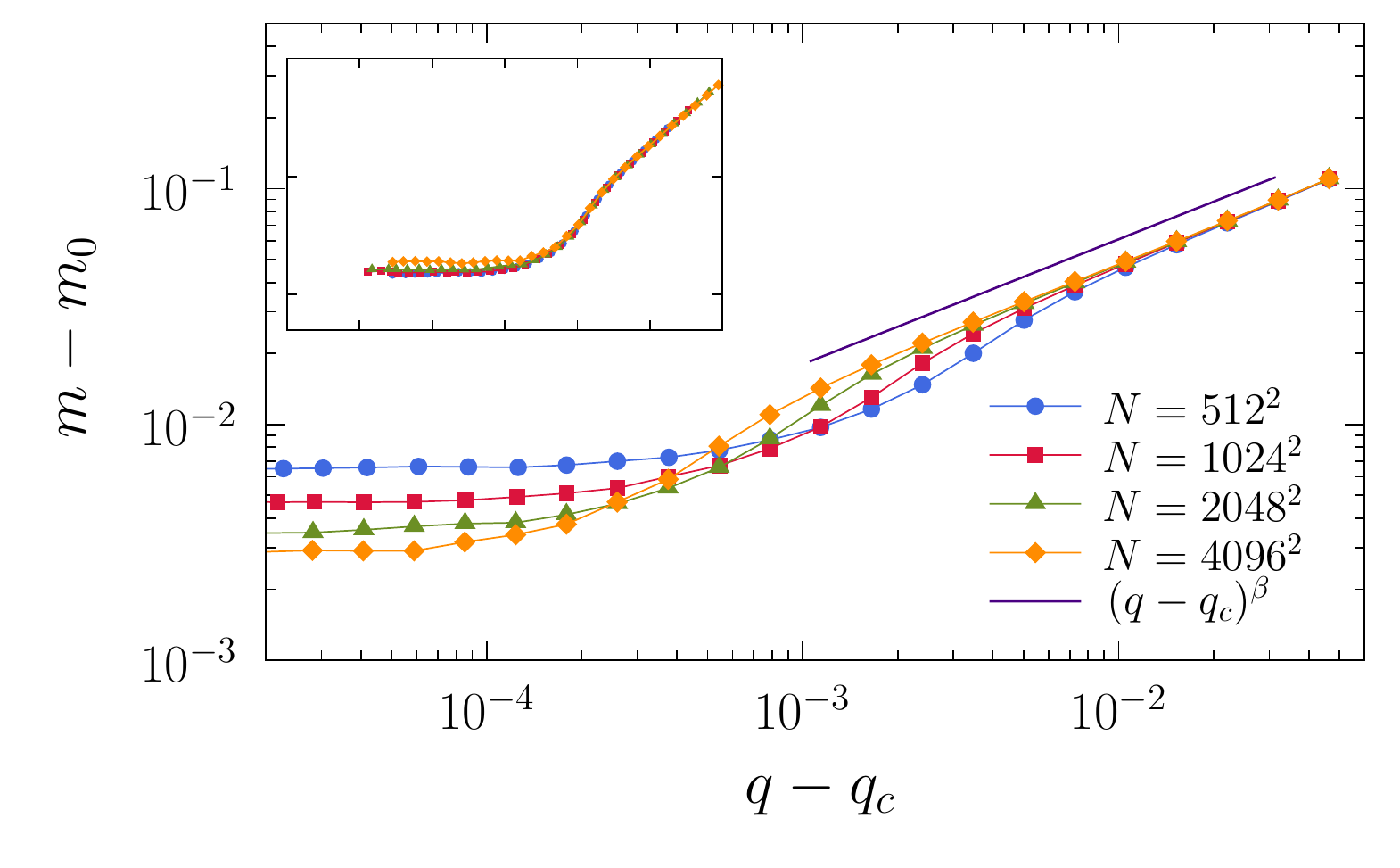}
\caption{(Color online)
$m-m_0$ vs $q-q_c$ is plotted for different system sizes on double logarithmic scales. The data seem to collapse into a single line of slope $\beta_m \approx 0.53$ in the large-$\Delta q$ region. Inset: Plot of the rescaled order parameter $(m-m_0)N^{\beta_m/\nbar_m}$ vs $\Delta q N^{1/\nbar_m}$. In order to achieve data collapse, we had to use $\nbar_m \approx 2.1$ and $\beta_m \approx 0.53$.
}
\label{fig:2dorder}
\end{figure}

The method of \cref{sec:beta-proof} can be used to numerically calculate the critical threshold $q_c$ and the jump size $m_0$. However, throughout this subsection, we will adopt the values $q_c=0.682892(5)$ and $m_0=0.603(2)$ which were recently obtained by Grassberger~\cite{grassberger}.

Theoretical consideration for the value of $\beta_m$ suggest $\beta_m=0.5$. \cref{fig:2dorder} shows a plot of $m(q)-m_0$ vs.\ $\Delta q \equiv q-q_c$ for various system sizes. In the not too small $\Delta q$ region, the data collapses into a single line, which enables us to measure $\beta_m \approx 0.53$. Since the region of agreement is quite short, we suspect that this deviation from the theoretical value is due to the finite size corrections. The scaling plot $(m-m_0)N^{\beta_m/\nbar_m}$ vs $\Delta q N^{1/\nbar_m}$ suggests $\nbar_m =2.1 \pm 0.2$.

\begin{figure}[h]
\includegraphics[width=.95\linewidth]{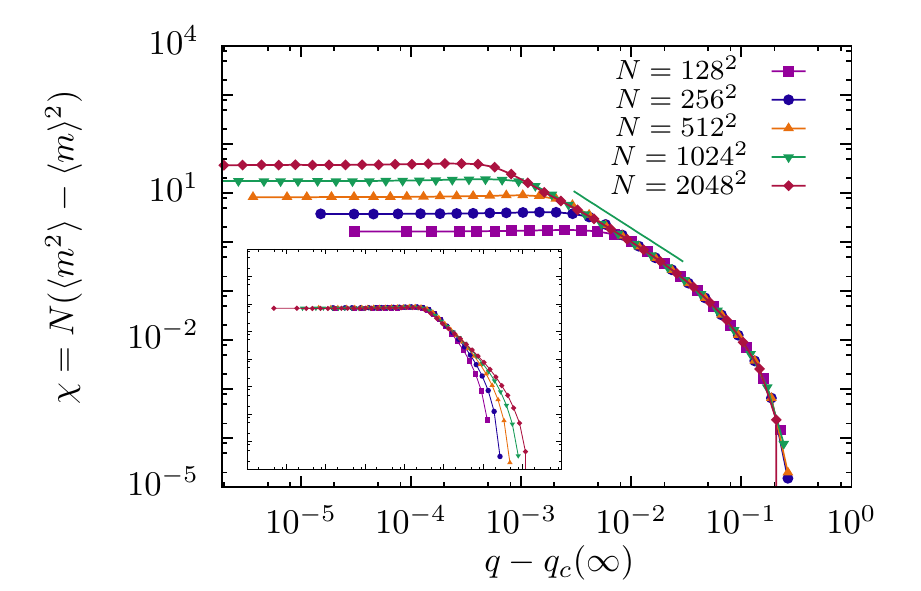}
\caption{(Color online)
Plot of the susceptibility $\chi\equiv N(\langle m^2 \rangle-\langle m\rangle^2)$ vs $q-q_c(\infty)$ for different system sizes on double logarithmic scales using systems with periodic boundary conditions. Inset: Data collapsed plot of ($\langle m^2 \rangle$-$\langle m \rangle^2$)$N^{1-\gamma_m/\nbar_m}$ vs $w\equiv \Delta q N^{1/\nbar_m}$ for the different system sizes. The best collapse is achieved using $\gamma_m\approx 1.35$ with $\nbar_m\approx 2.4$. The boundary conditions have a strong effect on the corrections to scaling and on the measured effective values of the exponents.
}
\label{fig:2dsusc}
\end{figure}

\cref{fig:2dsusc} shows the raw plot of the susceptibility $\chi = N(\langle m^2\rangle-\langle m\rangle^2)$ against $q-q_c$ for different system sizes. 
Due to strong corrections to scaling the exponent $\gamma$ is less accurate than for the ER case. The inset of \cref{fig:2dsusc} shows $(\langle m^2 \rangle-\langle m \rangle^2)N^{1-\gamma_m/\nbar_m}$ vs $\Delta q N^{1/\nbar_m}$ using $\gamma_m = 1.35 \pm 0.15$ and $\nbar_m = 2.4 \pm 0.2$, and one can observe that the deviation from power law leads to failure of collapse for large-$\Delta q$ region.

Our simulation data shows the following values of the exponents $\beta_m = 0.53 \pm 0.02$, $\gamma_m = 1.35 \pm 0.10$, and $\nbar_m = 2.2 \pm 0.2$. These exponents satisfy the scaling relation $\nbar_m = 2 \beta_m + \gamma_m$ within their error ranges.

\subsection{Critical behavior of avalanche dynamics}

We now examine the avalanche dynamics 2D lattices described above. Analogously to the case of double-layer ER networks, we denote the avalanche size at $q$ by $s_a(q)$, and the distribution of avalanche size by $\psa(q)$.

\begin{figure}[h]
\includegraphics[width=.95\linewidth]{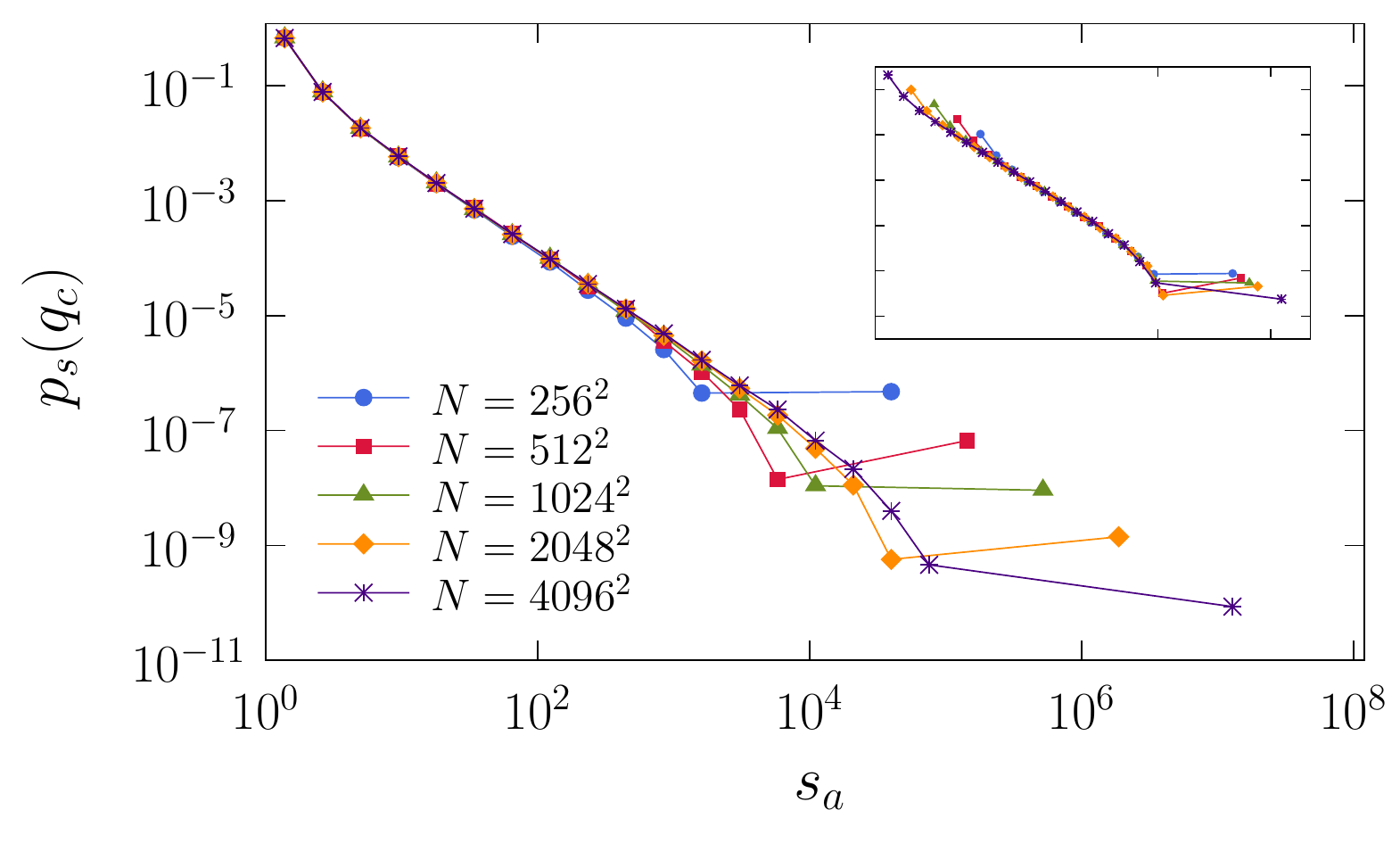}
\caption{(Color online)
Plot of the \asd{} $\psa(q_c)$ vs $s_a$ for different system sizes. The power-law regime is longer for larger system sizes.
Inset: Plot of the \asd{} in a scaling form $\psa(q_c) N^{\tau_a/\sigma_a \nbara}$ vs $s_aN^{-1/\sigma_a \nbara}$ for different system sizes. The data are well collapsed onto a single curve with $\sigma_a \nbara$.
}
\label{fig:2davalqc}
\end{figure}

The avalanche size distribution follows a power law at $q_c$ as $\psa(q_c) \sim s_a^{-\tau_a}$. The exponent is measured to be $\tau_a = 1.59 \pm 0.02$, see \cref{fig:2davalqc}. The avalanche size distribution follows this power-law up to a characteristic size $s_a^*$ that scales with the size of the system as $s_a^* \sim N^{1/\sigma_a\nbara}$, from which point it decays exponentially. The inset of \cref{fig:2davalqc} plots $\psa N^{\tau_a/\sigma_a\nbara}$ against $s_a N^{-1/\sigma_a \nbara}$ using $\sigma_a \nbara \approx 1.47$, with which the data collapses into a single line.

\begin{figure}[h]
\includegraphics[width=.95\linewidth]{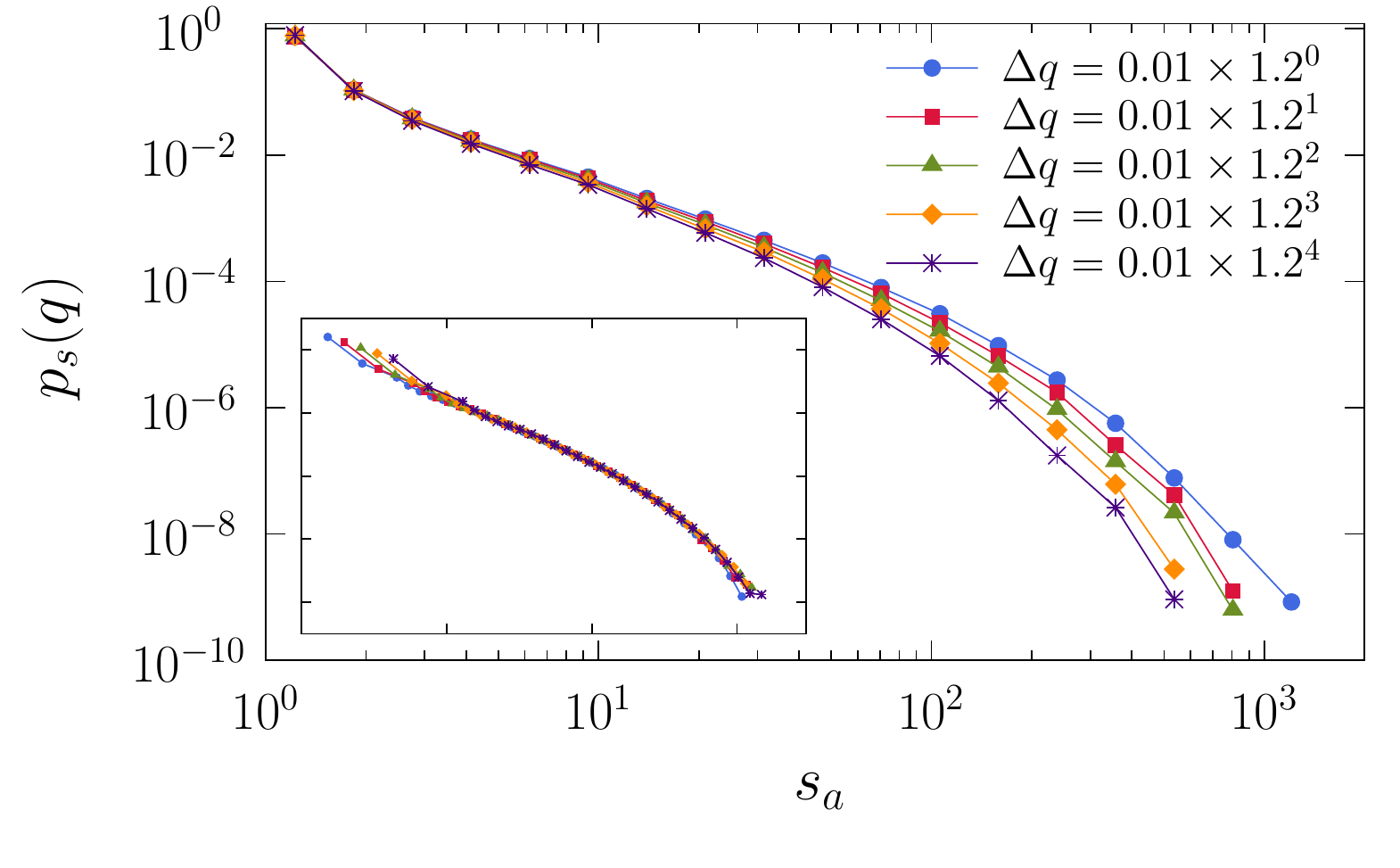}
\caption{(Color online)
Plot of $\psa(q)$ vs $s_a$ at different $\Delta q$ for $N=4096^2$. Inset: Plot of $\psa(q)/(\Delta q)^{-\tau_a/\sigma_a}$ vs $s_a(\Delta q)^{1/\sigma_a}$ at various $\Delta q$ for $N=4096^2$. With $\sigma_a \approx 0.70$, the data collapse into a single line.
}
\label{fig:2dpsa}
\end{figure}

\cref{fig:2dpsa} shows plots the \asd{} at various $q>q_c$ for a fixed system size $N=4096^2$. The distribution $\psa$ follows $\psa \sim s_a^{-\tau_a}f(-s_a/s_a^*)$, where $f$ is the so-called ``master curve'' (a scaling function) and we assume $s_a^* \sim (q-q_c)^{-1/\sigma_a}$. We obtain the exponent $\sigma_a$ by plotting $\psa(q)/(\Delta q)^{\tau_a/\sigma_a}$ versus $s_a(\Delta q)^{1/\sigma_a}$. The best data collapse is observed with $\sigma_a=0.70 \pm 0.05$, implying $\nbara = 2.1 \pm 0.2$, see \cref{fig:2dpsa}.
These values are confirmed by a somewhat more reliable method using the cumulative distribution function $P_s(q)$ which scales as $1-P_s(q)\sim s_a^{1-\tau_a}F(-s_a/s_a^*)$ where $F$ is another scaling function.

Notice in \cref{fig:2dpsa} that the cutoff sizes $s_a^*$ are small, and to increase them one has to carry out the measurement of the cascade size distribution close to the critical point. For this, a trade-off is to be made. Going too close to the critical point of the infinite system the critical behavior of the finite system is lost.

\begin{figure}[h]
\includegraphics[width=.95\linewidth]{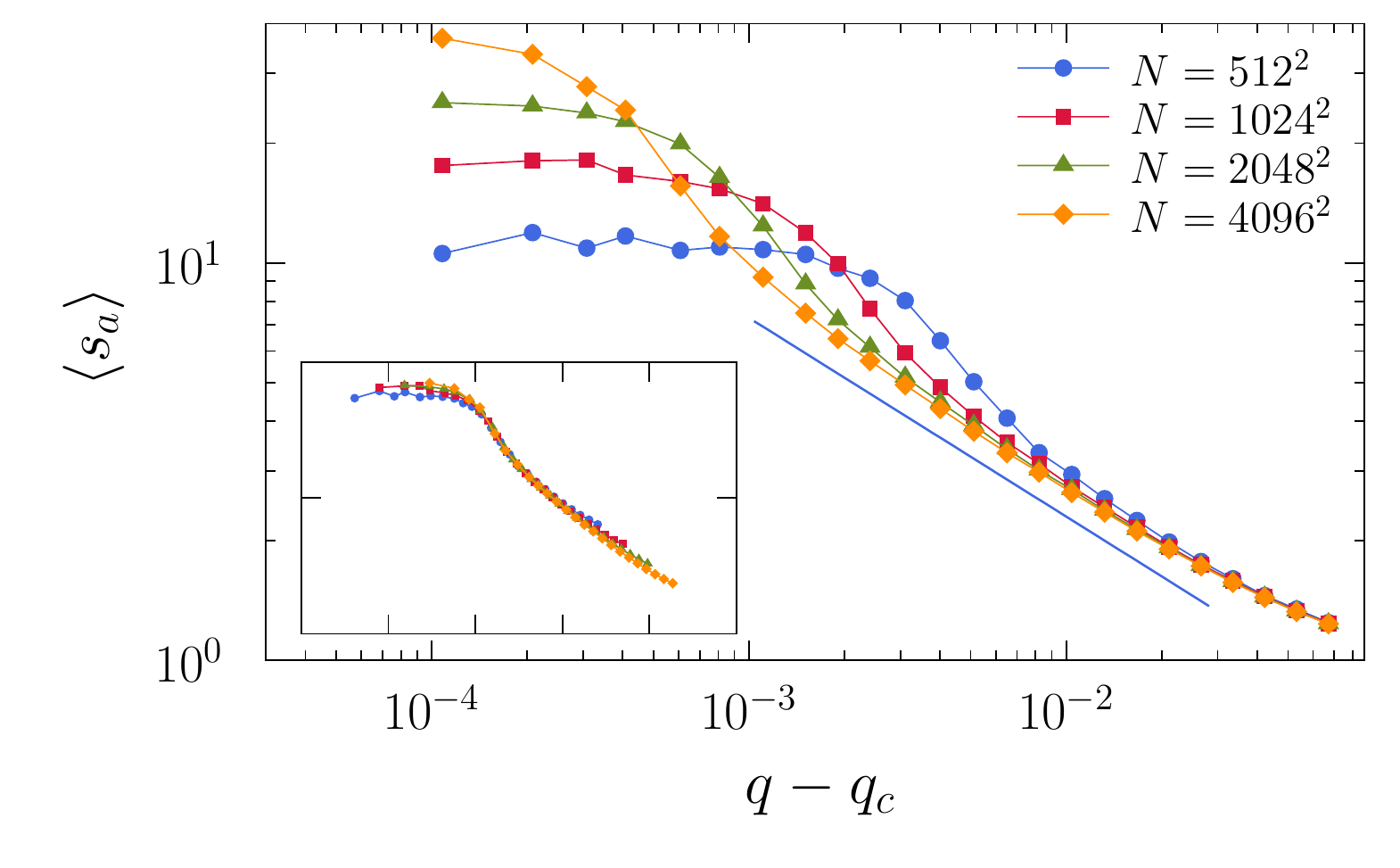}
\caption{(Color online)
The average size $\langle s_a \rangle$ of finite avalanches is plotted against $\Delta q$ for various system sizes. A line exhibiting the expected power-law with exponent $\gamma_m$ is drawn for comparison. The smaller systems seem to be too small to observe power-law behavior.
Inset: Plot of $\langle s_a \rangle N^{-\gamma_a/\nbara}$ vs $\Delta q N^{1/\nbara}$ for various system sizes. Using $\gamma_a \approx 0.5$ and $\nbara \approx 2.1$, the data roughly collapse in the mid-$\Delta q$ region. However, collapse fails in the large-$\Delta q$ region.
}
\label{fig:2d_avg_s}
\end{figure}

This observation supports the speculation that even systems as large as $N=4096^2$ are not enough to correctly assess power-law behaviors in the near-$q_c$ regions. As we shall see now, it is also related to the behavior of the first moment of the \asd. The first moment of $\psa$ follows $\langle s_a\rangle \equiv \sum_{s_a=1}'s_a\psa(q) \sim (q-q_c)^{-\gamma_a}$. \cref{fig:2d_avg_s} depicts our simulation results for the average size of finite avalanches for various $\Delta q$ and $N$, with a guideline of slope $\gamma_a = 0.5$ giving the best estimate for $\gamma_a$. The inset of \cref{fig:2d_avg_s} is a plot of $\langle s_a\rangle N^{-\gamma_a/\nbara}$ versus $\Delta q N^{1/\nbara}$ for different system sizes. Collapse is achieved with $\gamma_a \approx 0.50 \pm 0.05$ and $\nbara \approx 2.1 \pm 0.2$. This value of $\gamma_a$ reasonably satisfies the scaling relation $\gamma_a=(2-\tau_a)/\sigma_a$ within error ranges. The quality of the collapse is still unsatisfactory, which makes the values of these exponents questionable.

\subsection{Statistics of the number of hops}

\begin{figure}
\includegraphics[width=.95\linewidth]{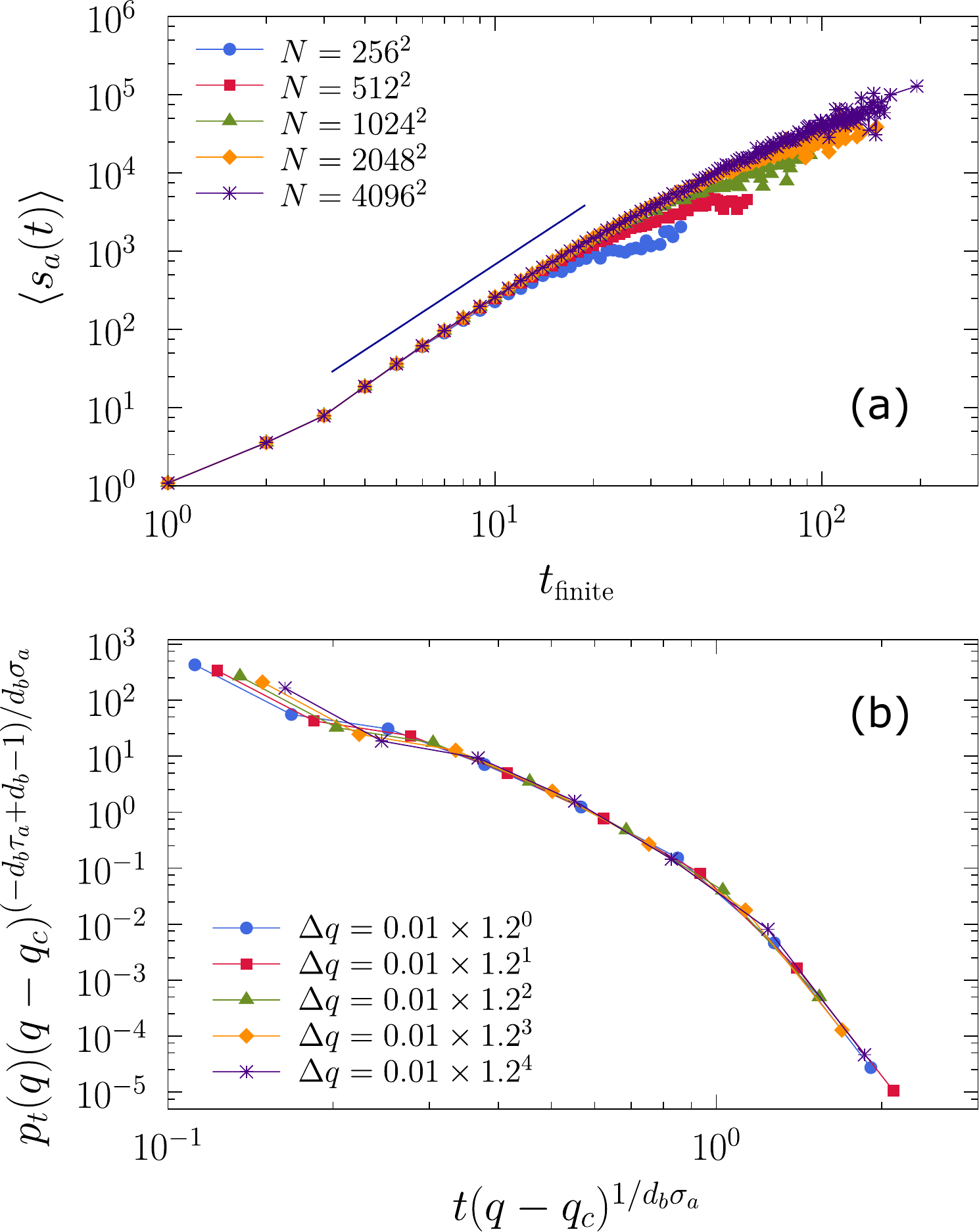}
\caption{(Color online)
(a) Plot of the mean avalanche sizes $\langle s_a \rangle$ vs the number of hops $t$ between the two layers. The overall slope is estimated to be about $2.75$.
(b) The distribution $p_t(q)$ of the number of hops $t$ vs $\Delta q$ in scaling form. $p_t(q)(\Delta q)^{(-d_b\tau_a+d_b -1)/d_b\sigma_a}$ is plotted against $t(\Delta q)^{1/d_b\sigma_a}$, with $d_b \approx 2.75$.
}
\label{fig:2dh}
\end{figure}

We now turn our attention to the number of hops $t$, starting with the hops in finite avalanche processes. One can see in \cref{fig:2dh}(a) that the average size of avalanches $\langle s_a \rangle_t$ roughly scales with $t$ as $\langle s_a \rangle_t \sim t^{2.75}$, meaning that the fractal dimension of avalanche trees is $d_b=2.75$, which is different from that of the case of ER networks. This allows us to assume that the characteristic number of hops $t^*$ roughly scales as $t^* \sim (s_a^*)^{1/d_b} \sim (q-q_c)^{(-1/d_b\sigma_a)}$. Then, the distribution of the number of hops for finite avalanches would satisfy $p_t(q) \sim t^{-d_b \tau_a + d_b -1}f(t^{d_b}/(\Delta q)^{-1/d_b\sigma_a})$. This behavior is confirmed by \cref{fig:2dh}(b), which shows a scaling plot of this distribution.

\begin{figure}
\includegraphics[width=.95\linewidth]{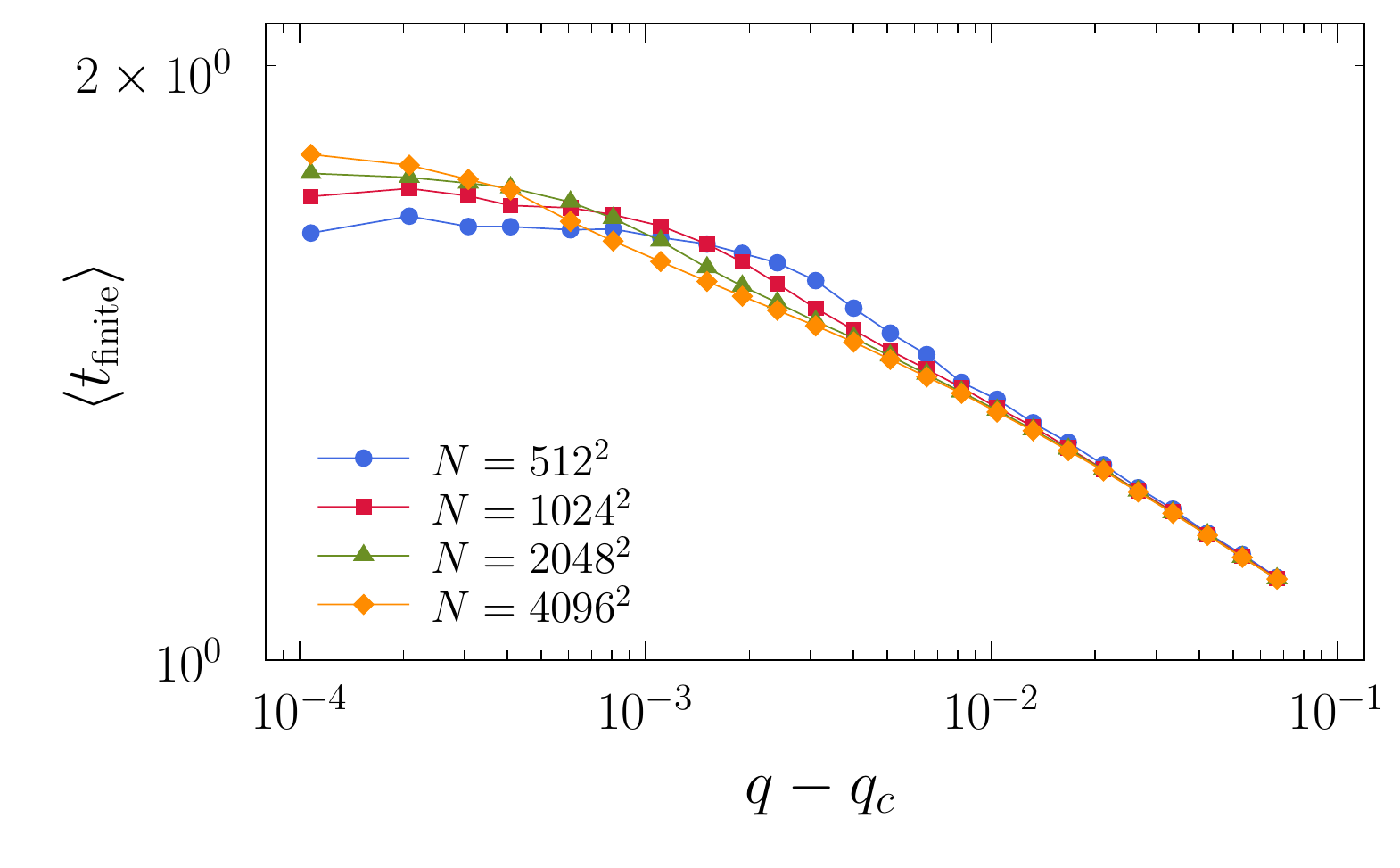}
\caption{
(Color online) The average number of finite hops $\langle t_\textrm{finite} \rangle$ is plotted against $\Delta q$ in double logarithmic scales.
}
\label{fig:2d_h_vs_Deltaq}
\end{figure}

\begin{figure}[h]
\includegraphics[width=.95\linewidth]{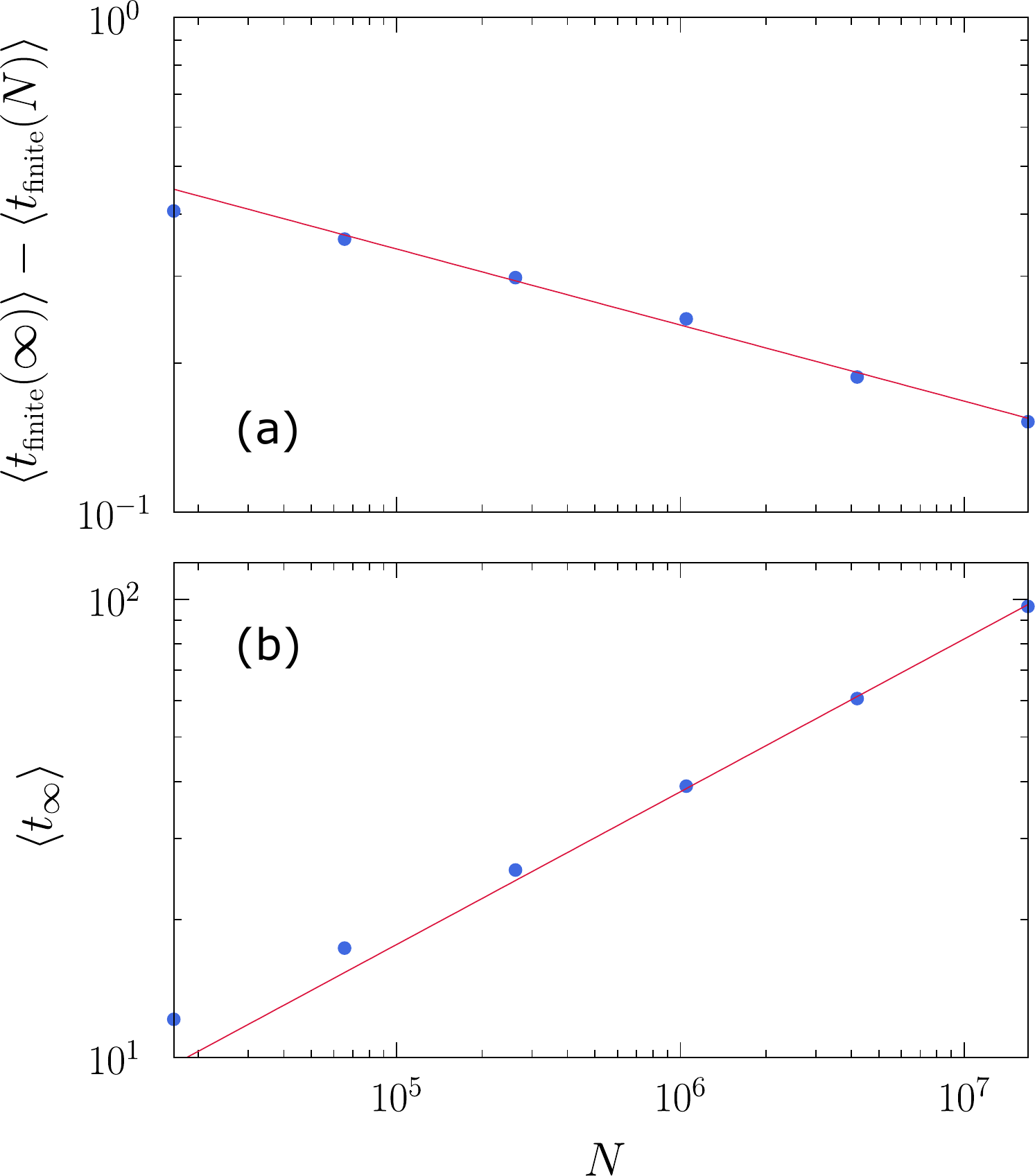}
\caption{
(Color online) (a) Plot of $\langle t_\textrm{finite}(\infty) \rangle - \langle t_\textrm{finite}(N) \rangle$ vs $N$ at $q_c$ in logarithmic scale, where $\langle t_\textrm{finite}(\infty) \rangle = 1.96$ was used.
(b) Plot of $\langle t_\infty \rangle$ of infinite avalanches vs $N$ in logarithmic scale. The guideline has a slope $0.33$.
}
\label{fig:2dh_at_pc}
\end{figure}

Recall that the value of $\tau_a$ was measured to be $\tau_a \approx 1.59$. This implies that, in contrast to the case of ER networks, the average number of hops of finite avalanches $\langle t \rangle \equiv \sum_{t=1}' t p_t(q)$ does not decrease logarithmically but rather follows a power-law with exponent $1-d_b\tau_a+d_b$. Also, the average number of finite hops $\langle t \rangle$ at $q=q_c$ approaches some value with a power-law, rather than increasing logarithmically
\cref{fig:2d_h_vs_Deltaq} and \cref{fig:2dh_at_pc}(a) illustrate these points.

Lastly, we consider the number of hops that constitute the infinite avalanches. Our simulation results reveal that this number scales as $N^{0.33}$. \cref{fig:2dh_at_pc}(b) shows these behaviors by plotting $\langle t_\infty \rangle$ against system size $N$.

In short, analogously to the two-layered ER network, the two sets of exponents $\{\beta_m, \gamma_m, \nbar_m \}$ and $\{\gamma_a, \nbar_a, \sigma_a, \tau_a \}$ are measured to be distinct. In this model too, values of the critical exponents measured through simulation satisfy the scaling relation $1-\beta_m=\gamma_a$ that relates these sets.
However, in all aspects the scaling behavior of 2D interdependent networks is much worse than that of the ER interdependent networks, indicating severe corrections to scaling.

\section{Analytic results}\label{sec:analytic}

In the following we derive two rules that hold for general interdependent networks.

\subsection{Proof of $\beta_m=1/2$ and $\gamma_a=1/2$}
\label{sec:beta-proof}
For the exponent $\beta_m$ values close to $1/2$ were measured for very different network settings~\cite{baxter}. We prove that $\beta_m=1/2$ holds for a wide range of mutual percolation processes.
Let $P_\infty^s(q)$ denote the fraction of nodes belonging to the giant component of the classical (single layer) percolation problem where $q$ is the fraction of occupied nodes. Let $q_c^s$ denote the critical point of this single layer percolation. If an additional layer is added to the percolation process with dependency links the critical point for the mutual percolation is $q_c\geq q_c^s$~\cite{buldyrev,bashan}.
Now let's consider a two-layered interdependent network with random infinite range interdependency links that represent a random one-to-one mapping between the layers. The control parameter $q$ denotes the fraction of the nodes kept. It has been shown that the size of the MCGC after the $i$th step is $P_\infty^s(q_i)$ where $q_i$ is an equivalent random attack given by the recursion~\cite{buldyrev}
\begin{equation}\label{eq:rec} q_i=\frac{q}{q_{i-1}} P_\infty^s(q_{i-1}). \end{equation}
The recursion has a fixed point $x(q)$ corresponding to the steady state $m(q)\equiv P_\infty^s(x(q))$ of the system:
\begin{equation}\label{eq:fixed} x^2=q P_\infty^s(x). \end{equation}
As $q_c > p_c^s$ the $P_\infty^s$ curve of single layer percolation can be approximated by its series near $q_c$:
\begin{equation}\label{eq:lin} P_\infty^s(q)= a+b\cdot q + O(q^2) \end{equation}
with $a\equiv P_\infty^s(q_c)$.

For the critical behavior close to $q_c$ we need to solve $ x^2(q) = q\cdot\big(a+bx(q)\big)$ resulting
\begin{equation}\label{eq:roots} x=\frac{bq\pm\sqrt{b^2q^2+4aq}}{2}. \end{equation}
At the critical point $q_c$ the determinant $b^2(q_c)^2+4aq_c$ is zero.
Introducing $q:=q_c+\Delta q$ and substituting into the valid (greater) result of~\cref{eq:roots} we have
\begin{equation}\label{eq:diff} x(q)-x(q_c)=\frac{b\Delta q+\sqrt{b^2(q_c+\Delta q)^2+4a(q_c+\Delta q)}}{2}. \end{equation}
By \eqref{eq:lin} and $b^2q_c^2+4aq_c=0$, we get
\begin{equation} 
m(q)-m(q_c)\sim (q-q_c)^{1/2}+ O(q-q_c). \end{equation}
Thus, we conclude $\beta_m=1/2$. Due to the sum rule (see next subsection) this also implies $\gamma_a=1/2$.

\subsection{Sum rule for interdependent networks and $\gamma_a=1-\beta_m$}\label{sec:sum-rule}

For the avalanche dynamics we summarize over the whole history of the network:
\begin{equation}\label{eq:newsum} 
1=m(q)+\int_q^1{\sum_s}^{\prime}{s p_s(\tilde{q})} \,\mathrm{d}\tilde{q}. 
\end{equation}
This formula expresses the fact that a site can either belong to the MCGC (first term on the r.h.s. of \cref{eq:newsum}) or it is eliminated in one of the avalanches (the sum in \cref{eq:newsum}).
Here $p_s(q)$ is the number of avalanches of size $s$ occurring per site per attack $\mathrm{d}q$ at $q$. The summation is carried out over all finite avalanches and the integral takes into account any events that were triggered for $\tilde{q}\in[q,1]$. 
It is useful to write \cref{eq:newsum} in differential form:
\begin{equation}
\frac{\mathrm{d}m(q)}{\mathrm{d}q}={\sum_{s}}^{\prime}{sp_s(q)}. 
\end{equation}
Since $m(q)-m(q_c)\propto (q-q_c)^\beta_m$, it yields $\mathrm{d} m(q)/\mathrm{d}q\propto (q-q_c)^{\beta_m-1}$. The right hand side describes the average size of finite avalanches which scales as $\langle s_a(q)\rangle\sim(q-q_c)^{-\gamma_a}$. Comparing the two sides we find that the relation $\gamma_a=1-\beta_m$ between the two set of exponents holds universally.

\begin{figure*}
\includegraphics[width=.9\linewidth]{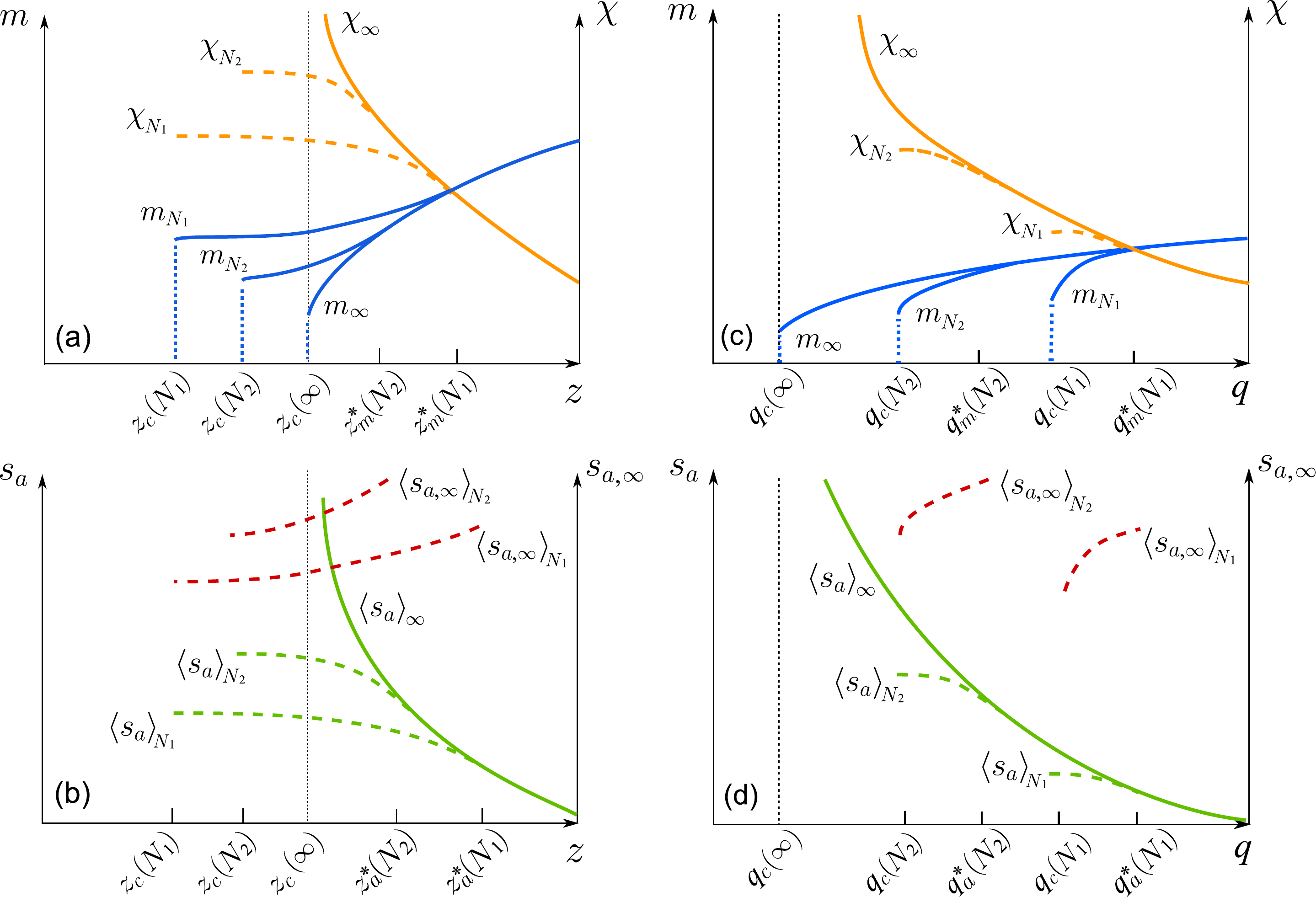}
\caption{(Color online) For Erd\H{o}s--R\'enyi (ER) interdependent networks, (a) schematic plot of the order parameter $m(z)$, the size of the giant MCC  per node (solid curves with dark gray (blue)) as a function of mean degree $z$, where $m(z)$ is averaged over surviving configurations with nonzero $m(z)$. Dotted lines represent the discontinuity of the order parameter. $z_{c}(N_i)$ ($i=1$ or 2) is the transition point obtained by averaging transition points over all runs, where $N_1 < N_2$. The susceptibility $\chi\equiv N(\langle m^2 \rangle - \langle m \rangle^2)$ (solid and dashed curves with light gray (orange)) is shown as a function of $z$. $z^{*}_{m}(N_i)$ is a crossover point across which finite-size effect appears in the side $ z < z_m^*$. Dashed curves in the interval [$z_{c}(N_i)$, $z^{*}_{m}(N_i)$]  represent the susceptibility in finite-size systems. 
(b) Schematic plot of the mean size of {\it finite avalanches} $\langle s_a \rangle_{N_i}$ for system size $N_i$ (solid and dashed curves with light gray (green)). Dashed curves represent $\langle s_a \rangle_{N_i}$ in finite-size systems, which occur for $z < z_a^*(N_i)$. $\langle s_{a,\infty} \rangle_{N_i}$ with 
dark gray (red) denotes mean size of {\it infinite avalanches} as a function of $z$. Here, the term ``infinite" 
refers to the avalanches that lead to complete collapse of the GMCC.
Note that $z_{m}^*(N_i)-z_c(\infty)$ and $z_{a}^*(N_i)-z_{c}(\infty)$ do not scale in the same manner with respect to $N$.
All four averages $m_{N_i}$, $\chi_{N_i}$, $\langle s_a \rangle_{N_i}$ and $\langle s_{a,\infty} \rangle_{N_i}$ are displayed only for $z>z_c(N_i)$.
(c) and (d) are similar schematic plots for the 2D interdependent networks with the semiperiodic boundary condition.
}
\label{fig:summary}
\end{figure*}

\section{Summary}\label{sec:summary}

Our aim has been in this paper to clarify the unusual features of the HPT as observed in the interdependent CF model. Due to the efficient algorithm~\cite{hwang} we were able to carry out large scale simulations for the ER and 2D interdependent networks and determine numerically the exponents and the finite size scaling functions.

The specific challanges related to the HPT for the interdependent CF model come from the fact that, in contrast to ordinary percolation, we have here two divergent length scales as the system approaches the transition point and, correspondingly, two sets of exponents.
The critical properties we obtained are schematically shown in Fig.17 for the Erd\H{o}s--R\'enyi (ER) and for the 2D interdependent networks. One set of exponents, $\{\beta_m, \gamma_m, \nbar_m \}$ is associated with the  order parameter and its related quantities, and the other set $\{\tau_a, \sigma_a, \gamma_a, \nbar_a \}$ is associated with the avalanche size distribution and its related ones. The subscripts $m$ and $a$ refer to the order parameter and avalanche dynamics respectively.
 
The numerically estimated values of the critical exponents for the ER and the 2D cases are listed in Table I.  They reveal the unconventional character of the transition: the exponents $\bar{\nu}_m$ and $\bar{\nu}_a$ and $\gamma_m$ and $\gamma_a$ are different from each other, respectively.
For the ER and 2D cases, the hyperscaling relation $\nbar_m=2\beta_m+\gamma_m$ holds even though data collapsing for the 2D case is not as satisfactory as for the ER case. The relation $\sigma_a \nbar_a=\tau_a$ does not hold (Sec.~\ref{sec:num_result_ER} and \ref{sec:num_result_2D}).

We showed analytically that the two sets of critical exponents are not completely independent of each other; they are coupled through the relation $m(z)+\int_z^{z_0} \langle s_a(z) \rangle \mathrm{d}z =1$, where $z_0$ is the mean degree at the beginning of cascading processes. This relation leads to $\mathrm{d}m(z)/\mathrm{d}z= \langle s_a(z)\rangle$ and yields $1-\beta_m=\gamma_a$.
We also showed that for random interdependence links $\beta_m =1/2$. Our numerical values support these relations.

We classified avalanches in the critical region as finite and infinite avalanches. When an infinite avalanche occurs, the GMCC completely collapses, and the system falls into an absorbing state. We found that the mean number of hopping steps denoted as $\langle t \rangle$ between the two layers in avalanche processes depends on the system size $N$ in different ways for the different types of avalanches:   $\langle t \rangle \sim \ln N$ for finite avalanches, and $\sim N^{1/3}$ for infinite avalanches on the ER interdependent network. This difference in the scaling again underlines the peculiarities of the HPT: The infinite avalanche give rise to $m_0$, while the finite ones contribute to the critical avalanche statistics.

Our results present a unified picture of HPT, however, there are still open questions for further research. The strong corrections to scaling, especially for the 2D case should be understood. We have realized that the boundary conditions have a strong impact on the corrections and one should persue the investigation along this line. A real challange is to understand how the hybrid transition can be properly treated with the method of the renormalization group. Furthermore, it would be very interesting to see how other hybrid transitions fit into the presented framework.

\begin{table*}
\caption{List of numerical values of the critical exponents for various cases. $m$ and $a$ in the second column mean the cases related to the order parameter and 
avalanche, respectively.}
\begin{center}
\setlength{\tabcolsep}{12pt}
{\renewcommand{\arraystretch}{1.5}
\begin{tabular}{ccccccc}
    \hline
    \hline
    
    & & $\beta$ & $\tau$ & $\sigma$ & $\gamma$ & $\bar{\nu}$ \\
    \hline
    
    \multicolumn{2}{c}{ordinary ER} & 1 & 1.5 & 0.5 & 1 & 3 \\
    
    \multirow{2}{*}{interdependent ER}
    & $m$ & $0.5 \pm 0.01$ & - & - & $1.05 \pm 0.05$ & $2.1 \pm 0.02$ \\
    \cline{2-2}
    & $a$ & - & $1.5 \pm 0.01$ & $1.0 \pm 0.01$ & $0.5 \pm 0.01$ & $1.85 \pm 0.02$ \\
    
    \multicolumn{2}{c}{ordinary 2D} & $0.139$ & $1.055$ & $ 0.286$ & $2.389$ & $2.667$\\
    
    \multirow{2}{*}{interdependent 2D}
    & $m$ & $0.53\pm 0.02$  & - & -  & $1.35 \pm 0.10$ & $2.2\pm 0.20$ \\
    \cline{2-2}
    & $a$ & - & $1.59 \pm 0.02$ & $0.70 \pm 0.05$ &$0.5\pm 0.05$ &$2.1\pm 0.2$ \\
    
    \hline
    \hline
\end{tabular}}
\end{center}
\end{table*}

\begin{acknowledgments}
This work was supported by National Research Foundation in Korea with the grant No.\ NRF-2014R1A3A2069005.
JK acknowledges support from EU FP7 FET Open Grant No. 317532, Multiplex. 
\end{acknowledgments}


\begin{thebibliography}{99}
\bibitem{model_comment} In the original model defined in \cite{buldyrev}, the control parameter is the fraction of nodes removed from one layer of the networks. However, an equivalent model in terms of the mean degree is introduced by \cite{son_grassberger}. We use the latter model in this paper.
\bibitem{mcc} MCC  is defined as the component to which a node belongs is connected directly or indirectly to all the other nodes in the same MCC in each layer's  network. The largest MCC of size $O(N)$ is called GMCC.
\end{thebibliography}




  
%
%


\end{document}